\documentclass[a4paper,11pt]{article}
\pdfoutput=1 %

\usepackage{jcappub} %

\usepackage[T1]{fontenc} %

\usepackage[utf8]{inputenc}
\usepackage{aas_macros}
\usepackage{amsfonts}
\usepackage{amsmath}
\usepackage{amssymb}
\usepackage{graphicx}
\usepackage{soul}
\usepackage[dvipsnames]{xcolor}
\usepackage{hyperref}
\hypersetup{colorlinks=true,allcolors=teal}

\newcommand{\Mh}{\ensuremath{h^{-1}M_{\odot}}}

\newcommand{\Mpch}{\ensuremath{h^{-1}{\rm Mpc}}}

\newcommand{\figref}[1]{figure~\ref{#1}}

\newcommand{\secref}[1]{section~\ref{#1}}

\newcommand{\be}{\begin{equation}}
\newcommand{\ee}{\end{equation}}

\newcommand{\red}[1]{{#1}}
\newcommand{\oldred}[1]{{#1}}

\title{\boldmath The evolving role of astrophysical modelling in dark matter halo relaxation response}

\author{Premvijay Velmani}
\author{and Aseem Paranjape}
\affiliation{Inter-University Centre for Astronomy \& Astrophysics,\\ Ganeshkhind, Post Bag 4, Pune 411007, India}

\emailAdd{premv@iucaa.in}
\emailAdd{aseem@iucaa.in}

\abstract{
We study the change in the radial distribution of dark matter within haloes in response to baryonic astrophysical processes in galaxies at different epochs, investigating the role of astrophysical modeling in cosmological hydrodynamic simulations in producing the response. We find that the linear quasi-adiabatic relaxation with additional dependence on the halo-centric distance 
provides a good description 
not only at $z=0$, but also at an earlier epoch ($z=1$) in the IllustrisTNG simulation suite, with parameters being more universal across a much larger variety of haloes at $z=1$ than at $z=0$. Through systematic analysis of a large collection of simulations from the CAMELS project, we find that the baryonic prescriptions for both AGN and stellar feedbacks have a strong influence on the relaxation response of the dark matter halo. In particular, only the parameters controlling the overall feedback energy flux have an effect on the relaxation response, while the wind speed and burstiness have negligible effect on the relaxation at a fixed amount of energy flux. However, the exact role of these parameters on the relaxation depends on the redshift. We also study the role of a variety of baryonic astrophysical processes through the EAGLE physics variation simulations. While these depict a similar picture regarding the importance of feedback effects, they also reveal that the gas equation of state has one of the strongest influences on the relaxation response, consistent with the expectation from self-similar analyses.}

\begin{document}
\maketitle
\flushbottom

\section{Introduction}
\label{sec:intro}
Advancements in computational cosmology have produced state-of-the-art hydrodynamical simulations of cosmological volumes with realistic galaxies, starting from tiny fluctuations in an otherwise uniform matter distribution (see, e.g., OWLS \citep{2010MNRAS.402.1536S}, Illustris \citep{2014MNRAS.445..175G}, FIRE \citep{2014MNRAS.445..581H}, EAGLE \citep{2015MNRAS.446..521S}, Horizon-AGN \citep[][]{2017MNRAS.467.4739K}, SIMBA \citep[][]{2019MNRAS.486.2827D}, IllustrisTNG \citep{2019ComAC...6....2N}). They include both baryonic and dark matter simulated by a hydrodynamical technique and an N-body technique respectively, with their relative fraction set to the cosmic \oldred{baryon fraction (obtained by CMB constraints for example)} in the initial distribution. As dark matter tends to accumulate into gravitationally collapsed structures called haloes, baryonic gas is allowed to cool and condense further to the centers of those haloes, forming galaxies. However, many sub-galactic astrophysical processes, such as star formation, are not resolved by these simulations, where the individual particle/cell is at least several thousand solar masses. Instead, they rely on subgrid prescriptions -- with parameters calibrated to a set of empirical observations. \oldred{This includes, for example in EAGLE, the $z=0$ black hole mass galaxy stellar mass relation and galaxy stellar mass function at $z=0.1$ among others,} that are yet to be fully understood.

Since dark matter haloes can also form in a gravity-only scenario, they have been extensively studied through N-body simulations in that setting \citep[see e.g.,][]{1997ApJ...490..493N}, ignoring all the ambiguities and additional computational costs of simulating baryonic astrophysics. Nevertheless, the impact of baryonic processes on the dark matter can be significant and must be accounted for before comparing against observations  \citep[see e.g.,][]{2010MNRAS.407..435A}.

The response of dark matter halo to galaxy formation has multiple aspects; we focus on the radial matter distribution relevant for mass profiles, rotation curves, etc. Early studies of individual haloes modelled this response as adiabatic relaxation of orbiting dark matter particles \citep[][]{osti6457593,1984MNRAS.211..753B,1986ApJ...301...27B,1987ApJ...318...15R}. However, this idealistic model rarely predicts the dark matter distribution within haloes in hydrodynamical cosmological simulations \citep[e.g.,][]{2004ApJ...616...16G,2010MNRAS.407..435A}. Moreover, the halo response was found to vary widely across haloes and in different simulations \citep[][]{2004ApJ...616...16G,2006PhRvD..74l3522G,2010MNRAS.402..776P,2010MNRAS.406..922T,2010MNRAS.405.2161D,2010MNRAS.407..435A,2011MNRAS.414..195T,2016MNRAS.461.2658D,2019A&A...622A.197A,2022MNRAS.511.3910F,2023Velmani&Paranjape}, leading to the development of various models of the response, some of which are direct extensions of adiabatic relaxation \citep[e.g.,][]{2004ApJ...616...16G,2006PhRvD..74l3522G,2010MNRAS.407..435A}.

In one such quasi-adiabatic model \cite{2010MNRAS.407..435A}, the relation between the relaxation ratio (change in the dark matter radius) and the mass ratio (change in the enclosed total mass) has been proposed to be quadratic. However, this relaxation relation tends to be more complex and varies widely across haloes in more realistic cosmological simulations \oldred{with feedback outflows from galaxies} \citep{2023Velmani&Paranjape}. We have shown in a previous work \citep{2023Velmani&Paranjape} that introducing an additional dependence on the halo-centric radius makes the relation linear across a variety of haloes in both IllustrisTNG and EAGLE simulations at the present epoch ($z=0$). In the first part of this work, we investigate those results concerning the relaxation response among early epoch halo populations.

Previous works have shown that feedback from various galactic processes strongly influences halo relaxation \citep{2011MNRAS.414..195T}. These \oldred{processes} primarily affect the offset in the relaxation relation, quantifying the excess relaxation experienced by the dark matter \citep{2023Velmani&Paranjape}. Our previous results \citep{2023Velmani&Paranjape} reflect this through small but significant deviations in the relaxation offset across haloes with different star formation activities. Other baryonic processes in the galaxy may also play a role in halo relaxation. For example, the gas equation of state has been found to strongly affect the relaxation relation even in more tractable self-similar galaxy-halo systems \citep{2024VelmaniParanjape}. In this work, we systematically study these effects using simulations with variations in the feedback implementation.

The goal of this paper is to further understand the role of various astrophysical processes in the galaxy, such as feedback, in mediating the relaxation response of the halo. We study the relaxation in the CAMELS and EAGLE suite of simulations with variations in the baryonic subgrid prescription. We also explore the relaxation at an earlier epoch in the higher resolution IllustrisTNG simulation. Even with a particular baryonic prescription, the nature and strengths of the galactic processes are different from those at the present epoch ($z=0$).

This paper is organized as follows. We briefly describe the cosmological simulations employed in this work in \secref{sec:sims}, and we report our techniques in this investigation of the halo relaxation response from the simulations in \secref{sec:methods}. The main results of this work are presented in \secref{sec:results}, followed by the conclusion in \secref{sec:conclusion}.

\section{Simulations}
\label{sec:sims}
Here we describe the cosmological simulations employed in this work; these are from three different publicly available suites namely IllustrisTNG, EAGLE and CAMELS simulations.
\subsection{IllustrisTNG}
\label{sec:sims-IllTNG}
The IllustrisTNG simulations, conducted by the TNG collaboration, employed the \textsc{arepo} code \citep[][]{2020ApJS..248...32W}, which utilizes a moving mesh approach defined by Voronoi tessellation \citep[][]{2010MNRAS.401..791S}. These simulations incorporate an updated model of galaxy formation that includes cosmic magnetic fields in addition to major baryonic processes such as cooling, star formation, and stellar and AGN feedback \citep[][]{2017MNRAS.465.3291W,2018MNRAS.473.4077P}. The suite comprises three cosmological boxes: TNG50, TNG100, and TNG300, with periodic \oldred{comoving} box sizes of $35 \Mpch$, $75 \Mpch$, and $200 \Mpch$, respectively, consistent with the cosmology from \oldred{Planck results} \cite{2016A&A...594A..13P}. Initial conditions were generated at $z=127$ using the Zel'dovich approximation \citep[][]{1970A&A.....5...84Z} with the \textsc{N-GenIC} code \citep[][]{2015ascl.soft02003S}. 

We utilize the highest resolution runs from all three boxes to study a wide range of halo responses to galaxy formation. Specifically, TNG50 offers sufficient resolution for low-mass haloes, while TNG300 provides an adequate number of cluster-scale haloes. Throughout our analysis, we utilize data from redshifts $z=0.01$ and $z=1$ from IllustrisTNG for both hydrodynamical and corresponding gravity-only runs. This allows us to examine the effects of baryonic processes on dark matter halo properties across different scales and epochs.

\subsection{EAGLE}
\label{sec:sims-EAGLE}
The EAGLE (Evolution and Assembly of GaLaxies and their Environments) cosmological simulations were conducted using a modified version of the \textsc{gadget-3} code, which employs smoothed particle hydrodynamics \citep[][]{2005MNRAS.364.1105S}\oldred{, following the same cosmology as IllustrisTNG \cite{2016A&A...594A..13P}}. Initial conditions were generated using the \textsc{ic\_2lpt\_gen} code following \cite{2010MNRAS.403.1859J}. The main large-volume simulation was performed in a cosmological volume of $(100 ~\rm{Mpc})^3$ periodic box with its reference model of galaxy formation, incorporating sub-grid prescriptions for various baryonic processes such as cooling, star formation, and feedback mechanisms \citep[][]{2015MNRAS.446..521S,2015MNRAS.450.1937C}. This reference model of EAGLE simulations has been shown to produce realistic galaxies \citep[][]{2015MNRAS.448.2941S,2015MNRAS.450.4486F,2015MNRAS.452.2879T}. 

In addition, this suite includes multiple small-volume simulations with wide variations in the baryonic subgrid prescriptions for astrophysical processes. These simulations were performed at the same resolution as the large-volume reference simulation but in a $25 ~\rm{Mpc}$ periodic box. The variations include adjustments to the gas equation of state, the threshold for star formation, efficiency of stellar feedback, the viscosity of the black-hole accretion disk and the nature of stochastic heating caused by AGN feedback. In particular we study these following simulations:

\begin{itemize}
    \item \textbf{Ref} (Reference model): This simulation uses the standard EAGLE subgrid physics parameters but in this smaller box.
    \item \textbf{eos53} : This considers a gas equation of state with adiabatic index $\gamma = 5/3$.
    \item \textbf{eos1} : This considers an isothermal gas equation of state with $\gamma = 1$.
    \item \textbf{FixedSfThresh}: This uses a constant threshold \oldred{in gas density} for star formation independent of the metallicity.
    \item \textbf{WeakFB}: This follows a lower efficiency of stellar feedback, \oldred{with the asymptotic values $f_{\rm{th,min}}$ and $f_{\rm{th,max}}$} scaled down by $50 \%$ compared to the reference simulation. \oldred{The actual energy output per unit stellar mass formed is proportional to $f_{\rm{th}}$ which takes on values between these asymptotic values based on metallicity $(Z)$ and density $(n_{\rm{H,birth}})$ in the reference feedback model named $FBZ\rho$ as follows \citep{2015MNRAS.450.1937C,2015MNRAS.446..521S}:
    \begin{equation}
        f_{\rm{th}} = f_{\rm{th,min}} + \frac{f_{\rm{th,max}} - f_{\rm{th,min}} }{ 1 + \left( \frac{Z}{0.1 Z_{\odot}} \right)^{n_{Z}} \left( \frac{n_{\rm{H,birth}}}{n_{\rm{H,0}}} \right)^{-n_{n}} }
    \end{equation}
    where $n_{n} = n_{Z} = 2/\ln 10$, $Z_{\odot}=0.0127$ and $n_{\rm{H,0}}=0.67 \rm{cm}^{-3}$.}
    \item \textbf{StrongFB}: This follows a higher efficiency of stellar feedback, \oldred{with the asymptotic values $f_{\rm{th,min}}$ and $f_{\rm{th,max}}$} twice that of the reference simulation.
    \item \textbf{NoAGN}: AGN feedback is disabled in this simulation to study the effects of stellar feedback alone.
    \item \textbf{AGNdT8}: In this, the temperature of gas raised by the AGN feedback heating is lower with $\Delta T_{\rm{AGN}} = 10^8$K from the reference value of $\Delta T_{\rm{AGN}} = 10^{8.5}$K.
    \item \textbf{AGNdT9}: In this, the temperature of gas raised by the AGN feedback heating is higher with $\Delta T_{\rm{AGN}} = 10^9$K from the reference value of $\Delta T_{\rm{AGN}} = 10^{8.5}$K.
    \item \textbf{RefHR}: This simulation uses the same EAGLE subgrid prescription as `Ref' simulation but at a mass resolution higher by factor 8.
    \item \textbf{RecalHR}: This simulation uses the same resolution as `RefHR' but with recalibrated EAGLE subgrid prescription. \oldred{Particularly the feedback-related parameters have been modified to produce consistent $z\sim 0$ galaxy stellar mass function (GSMF) from the previously calibrated values in low resolution `Ref' simulation}.
\end{itemize}
We use the redshift $z=0$ data from this set of simulations along with their corresponding gravity-only run to study the role of different astrophysical processes on the relaxation response of the halo.

\subsection{CAMELS}
\label{sec:sims-CAMELS}
The Cosmology and Astrophysics with MachinE Learning Simulations (CAMELS) project comprises a comprehensive suite of hydrodynamical simulations designed to explore the interplay between cosmological and astrophysical parameters in shaping the universe's large-scale structure \cite[][]{CAMELS_presentation}. This large collection of simulations are performed in a relatively smaller cosmological volume of $(25 \ \mathrm{Mpc}/h)^3$, containing $256^3$ dark matter particles and an equivalent number of baryonic particles \cite{CAMELS_DR1}.\footnote{In addition, simulations in larger volume of $(50 \ \Mpch)^3$ are also expected to be made available as part of CAMELS project. It will be interesting to see the implications of improved statistics and large scale effects in a future work.} 

In our study, we specifically utilize the CAMELS-TNG suite's 1P set of simulations, a subset that methodically varies one parameter at a time to isolate the effects of individual parameters in the TNG model. For our analysis, we concentrate on astrophysical parameters related to supernova (SN) feedback and active galactic nucleus (AGN) feedback, each governed by the following two distinct parameters:
\begin{itemize}
    \item \textbf{Supernova Feedback Parameters:}
    \begin{itemize}
        \item $A_{\mathrm{SN1}}$: Varied between 0.25 to 4, this parameter controls the energy flux of the galactic winds. It is implemented as a prefactor for the overall energy output per unit star formation rate \cite{2018MNRAS.473.4077P,CAMELS_presentation}.
        \item $A_{\mathrm{SN2}}$: Varied between 0.5 to 2, this parameter controls the speed of the galactic winds. For a fixed $A_{\mathrm{SN1}}$, changes in $A_{\mathrm{SN2}}$ affect the galactic wind speed in concert with the mass-loading factor to maintain a fixed energy output.
    \end{itemize}
    \item \textbf{AGN Feedback Parameters:}
    \begin{itemize}
        \item $A_{\mathrm{AGN1}}$: Varied between 0.25 to 4, this parameter controls the overall power injected in the kinetic feedback mode of AGN. It is implemented as a prefactor for the energy per unit black-hole accretion rate \cite{2017MNRAS.465.3291W,CAMELS_presentation}.
        \item $A_{\mathrm{AGN2}}$: Varied between 0.5 to 2, this parameter controls the burstiness and the temperature of the heated gas during AGN feedback "bursts" by changing the wind speed of the AGN feedback.
    \end{itemize}
\end{itemize}
All these parameters have a value of unity in the fiducial simulation, and there are five simulations with higher values and another five simulations with lower values for each of these parameters. In total, we utilize these 41 hydrodynamical simulations and compare them against the gravity-only simulation over the same cosmological volume. \oldred{All these simulations follow the same cosmology, $\Omega_m=0.3, \Omega_b = 0.049, \Omega_k=0, \sigma_8 =0.8, n_s= 0.9624, h=0.6711$. Since these values are very close to the Planck constraints used in IllustrisTNG and EAGLE, we compare the relaxation results qualitatively between these simulations ignoring the effect of these differences in the cosmology.}

\section{Techniques}
\label{sec:methods}
\subsection{Haloes with relaxation}
\label{sec:hals}
In the simulations from IllustrisTNG and EAGLE suites, 3D friend-of-friends (FoF) algorithm \citep[see][for details]{2016A&C....15...72M,2019ComAC...6....2N} was used to obtain halo group catalogues, while the \textsc{subfind} code \citep{2001MNRAS.328..726S} was used to identify the subhaloes within these FoF group haloes as gravitationally bound substructures. Within each FoF group halo, the subhalo enclosing the gravitational potential minimum is assigned as its central subhalo. This trough in the gravitational potential is used to define the centre of that halo. Its size is characterized by the `virial' radius $R_{\rm vir}\equiv R_{\rm 200c}$ defined as the radius of the sphere around its centre enclosing a mean matter density that is 200 times the cosmological critical density. Its mass is then defined as the corresponding total mass enclosed $M\equiv M_{200c}$. In the simulations from the CAMELS suite, the haloes were identified in the phase-space by a 6D FoF algorithm using the \textsc{rockstar} code. \red{We chose ROCKSTAR because it has been shown to provide  more accurate halo positions compared to 3D FoF + \textsc{subfind} \cite{2013ROCKSTAR}, which is crucial for computing the relaxation response in the radial distribution of dark matter.} We characterize sizes and masses using the same quantities  $R_{\rm 200c}$ and $M_{\rm 200c}$ respectively. \red{For the IllustrisTNG main simulations, we continue to use FoF and \textsc{subfind}, but we plan to consider ROCKSTAR for these simulations in the future. This distinction could potentially lead to additional differences in the results, which we aim to explore in future analyses with ROCKSTAR halo catalogs of IllustrisTNG.}

To study the relaxation response of dark matter, we match the haloes from the full hydrodynamic simulations with the haloes in the corresponding gravity-only runs performed over same cosmological volumes. We identify these matched halo pairs based on the amount of overlap in their proto-haloes characterized by the positions of their particles in the initial conditions. In particular, this involves identifying nearby haloes of similar sizes between the hydrodynamical and gravity-only simulations \oldred{by organising the haloes in k-d tree structures. Among the nearby haloes, we measure the overlap in their respective proto-haloes to identify only the strongly matched haloes. For a given pair of hydrodynamic halo with one of the nearby gravity-only halo, we calculate the fraction of all particles in the hydrodynamic halo that has originated from the same location in the initial condition of the box as the gravity-only simulated halo. We then remove all those pairs with a matching fraction of less than $50\%$ between any one of them with respect to the other halo in the pair. This algorithm allows us to find matched halo for more than $95\%$ of the haloes of our interests} \citep[see][for details]{2023Velmani&Paranjape}.

\subsection{Relaxation Response Modeling}
\label{sec:methods-relmodel}
In the catalogues of matched halo pairs, hydrodynamical ones with galaxies are considered relaxed, while the gravity-only counterparts represent unrelaxed dark haloes. The relaxation response of dark matter within a halo, is generally evaluated through variations in their radial mass profiles indicating contraction or expansion due to the galaxy.  These spherically averaged dark matter profiles are computed by the cumulative sum of the mass contributed by all dark matter particles within concentric spherical shells. For the gravity-only halo, the cosmic dark matter fraction of the mass in each particle is considered as contributing to the dark matter. 

We employ the quasi-adiabatic relaxation framework to characterize the relaxation from these profiles.
The relaxation response in cold dark matter occurs entirely due to gravitational interactions with baryons. This is an aggregate effect of the baryonic mass flow resulting from galactic processes such as inflows and feedback. The quasi-adiabatic relaxation model is a physically motivated framework that relates the change in the spherically averaged dark matter distribution at a given time to the spherically averaged baryonic distribution at the same time. This baryonic profile encompasses all non-dark matter mass, including gas and stars.

Earlier models assumed spherical halo where the dark matter particles maintain their radial ordering while responding adiabatically to baryonic particle flows \citep[][]{1986ApJ...301...27B}. In this scenario, if a dark matter particle initially orbiting at radius \( r_i \) in the unrelaxed halo moves to radius \( r_f \) in the relaxed halo, the enclosed dark matter mass within these radii remains equal.
\begin{equation} 
M_f^d(r_f) = M_i^d(r_i)\,.
\label{eq:DMmass}
\end{equation}

However, due to the baryonic mass flow, the total mass enclosed within these spheres is not necessarily equal, \( M_i(r_i) \neq M_f(r_f) \). Further assumption of angular momentum conservation for dark matter particles in circular orbits, implies that the change in total enclosed mass must be consistent with the amount of relaxation \citep[][]{1986ApJ...301...27B}.
\begin{align}
    r_i \,M_i(r_i) = r_f \,M_f(r_f) %
    \implies 
    \frac{r_f}{r_i} = \frac{M_i(r_i)}{M_f(r_f)}\,. 
\label{eq:AR}
\end{align}
The \textbf{Quasi-adiabatic relaxation framework} empirically extends this idealized scenario by considering the relaxation ratio \( r_f/r_i \) as a function of the mass ratio \( M_i/M_f \).
\begin{align}
\frac{r_f}{r_i} &= 1 + \chi \left( \frac{M_i(r_i)}{M_f(r_f)} \right) 
\label{eq:qAR}
\end{align}
Various quasi-adiabatic models have been proposed, offering different approaches to this framework \citep{2010MNRAS.407..435A,2004ApJ...616...16G,2023Velmani&Paranjape}. A minimal extension considered in some of the baryonification procedures \citep[]{2015JCAP...12..049S,2021MNRAS.503.4147P}
incorporate dark matter response as a quasi-adiabatic relaxation with a single parameter
\be
\chi(y) = q\,(y-1)\,.
\label{eq:chi-linear}
\ee
Velmani \& Paranjape (2023) \cite{2023Velmani&Paranjape} proposed a locally linear model of the relaxation relation as follows:
\begin{align}
\label{eq:chi-linear-q0}
\frac{r_f}{r_i} - 1 &= q_1(r_f) \left[ \frac{M_i(r_i)}{M_f(r_f)} - 1 \right] + q_0(r_f),.
\end{align}
In that model, the relaxation is described by a linear relation however there is an additional explicit dependance on the halo centric distance.

\section{Results}
\label{sec:results}
We begin by exploring the relaxation response at an earlier redshift in the IllustrisTNG reference simulation. This allows us to understand how the different galactic processes occurring at earlier times affect dark matter halo relaxation. Following this, we examine variations in the baryonic prescriptions in the CAMELS and EAGLE simulations.

\subsection{Early epoch in IllustrisTNG simulations}
\label{sec:res-itng-z01}

In this section, we investigate the relaxation response in the radial distribution of dark matter in the main simulations of IllustrisTNG simulations at an earlier redshift of $z = 1$ and compare it to the present redshift of $z = 0$. Haloes at both these redshifts were identified and matched between hydrodynamical and the corresponding gravity-only runs as described in \secref{sec:hals}. Additionally, we use the SubLink merger tree catalogues to trace the most massive progenitor haloes at $z=1$ of the haloes considered at $z=0$ \cite{2019ComAC...6....2N}. While present epoch haloes are sampled only by their masses at the present time, we consider three different methods for sampling the early epoch haloes. This results in four distinct sets of halo samples:

\begin{enumerate}
    \item $z=1$ haloes sampled by their masses at $z=1$.
    \item $z=1$ haloes sampled by the masses of their descendants at $z=0$. Note that not all haloes with a given mass at the present time have valid progenitors with the same mass at $z=1$.
    \item $z=1$ haloes sampled by their masses at $z=1$, but the mass bins are defined by the median masses of the most massive progenitors of the $z=0$ haloes.
    \item $z=0$ haloes sampled by their masses at $z=0$.
\end{enumerate}

\begin{figure}[htbp]
\centering
\includegraphics[width=0.49\linewidth]{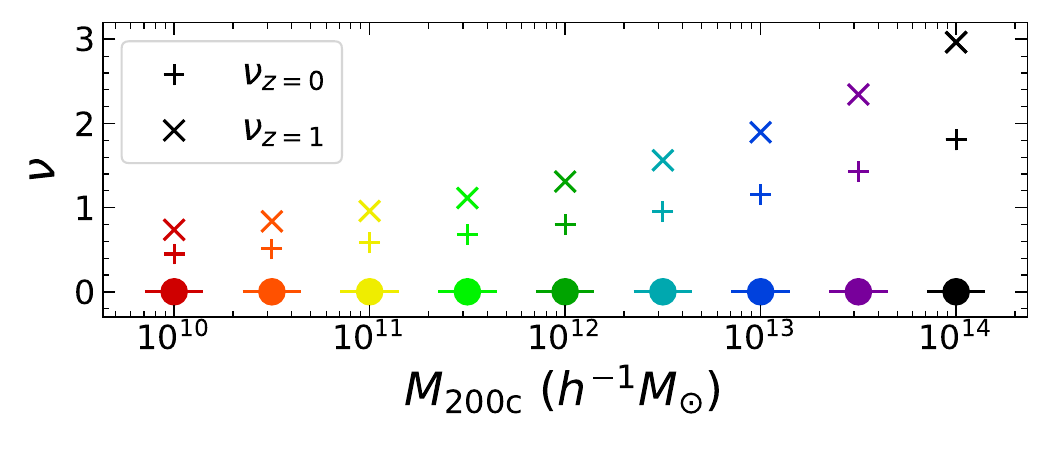}
\caption{Representative colors denoting each of the halo mass bins \oldred{with y-axis denoting} the corresponding values of peak heights $\nu$ at redshifts $z=0$ and $z=1$.}
\label{fig:mass_bin_label-z01}
\end{figure}

In each case, we consider the nine mass bins starting from $\log (M/\Mh) = 10$ to $14$ in steps of $0.5$ dex represented by the colors shown in \figref{fig:mass_bin_label-z01}. \oldred{We choose a conservative lower cut and ignore haloes of mass less than $10\Mh$ whose direct progenitors at $z=1$ have a radius of less than 100 times the smoothing length. And we also don't consider halo masses above $10^{14}\Mh$ with less than 50 haloes in similarly sized bins.} Additionally, this figure indicates the peak heights ($\nu$) corresponding to these halo masses at both $z=0$ and $z=1$. These values correspond to the rarity of haloes with that mass at that redshift, with rarer haloes having larger values of $\nu$. This can be used to identify the mass of the halo at $z=1$ that will have similar
rarity as $z=0$ halo of a given mass. 

\oldred{We study haloes with masses between $10^{10} \Mh$ \& $10^{12} \Mh$ from the smallest box TNG50, masses between $10^{11} \Mh$ \& $10^{13} \Mh$ from TNG100 and masses between $10^{12} \Mh$ \& $10^{14} \Mh$ from TNG300. We present the results from all three boxes together and discuss the convergence by comparing the results for some of those nine mass bins that were studied in more than one TNG box. In all our analysis, we ignore the region within ten times the smoothing length from the halo centre. Hence, for a given halo mass, the smallest box with the highest resolution allows us to probe the relaxation from the innermost regions, albeit with a smaller sample of haloes.}

\begin{figure*}
\centering
\includegraphics[width=0.48\linewidth,trim={0.5cm 0.5cm 0 0.5cm},clip]{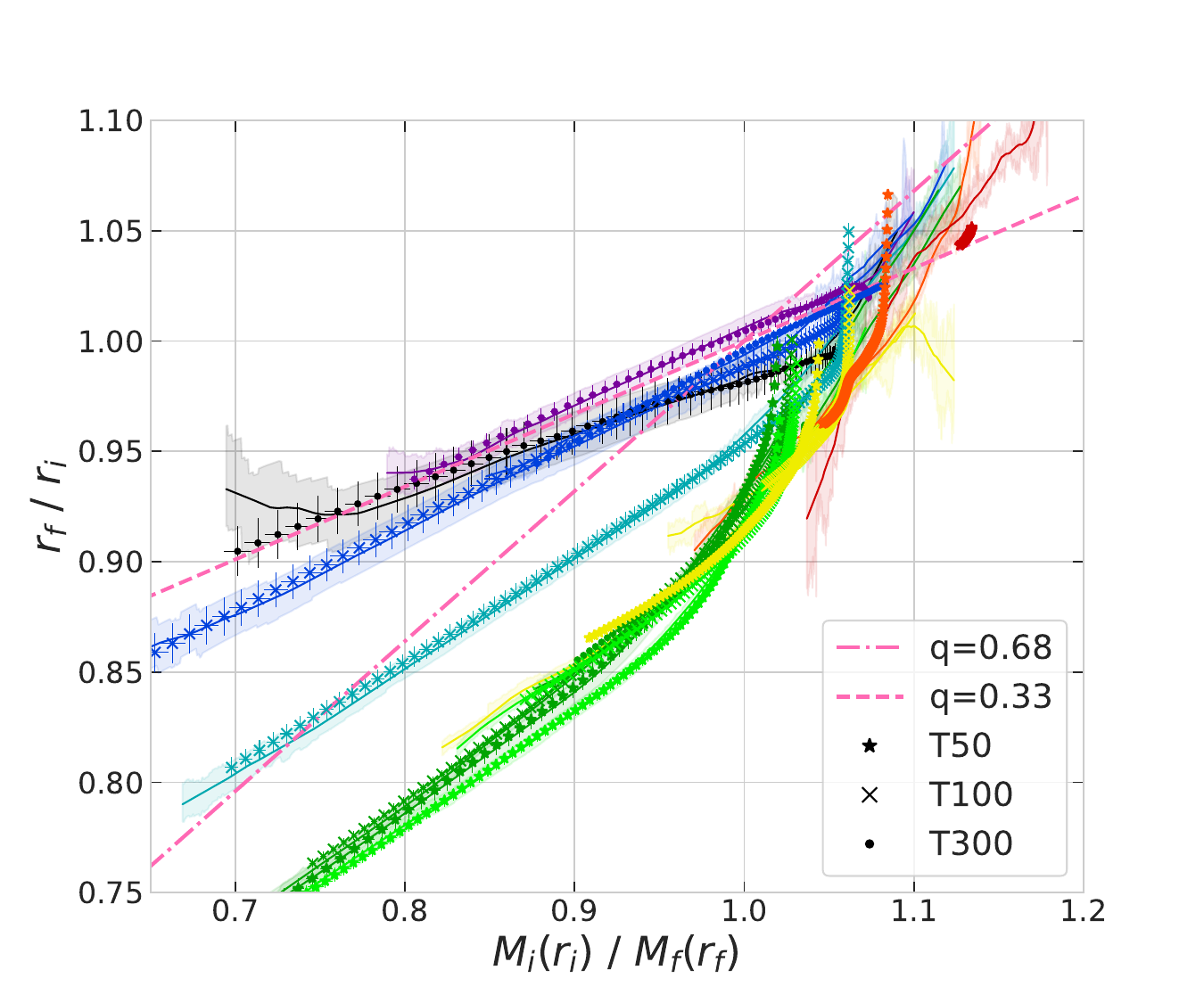}
\includegraphics[width=0.48\linewidth,trim={0.5cm 0.5cm 0 0.5cm},clip]{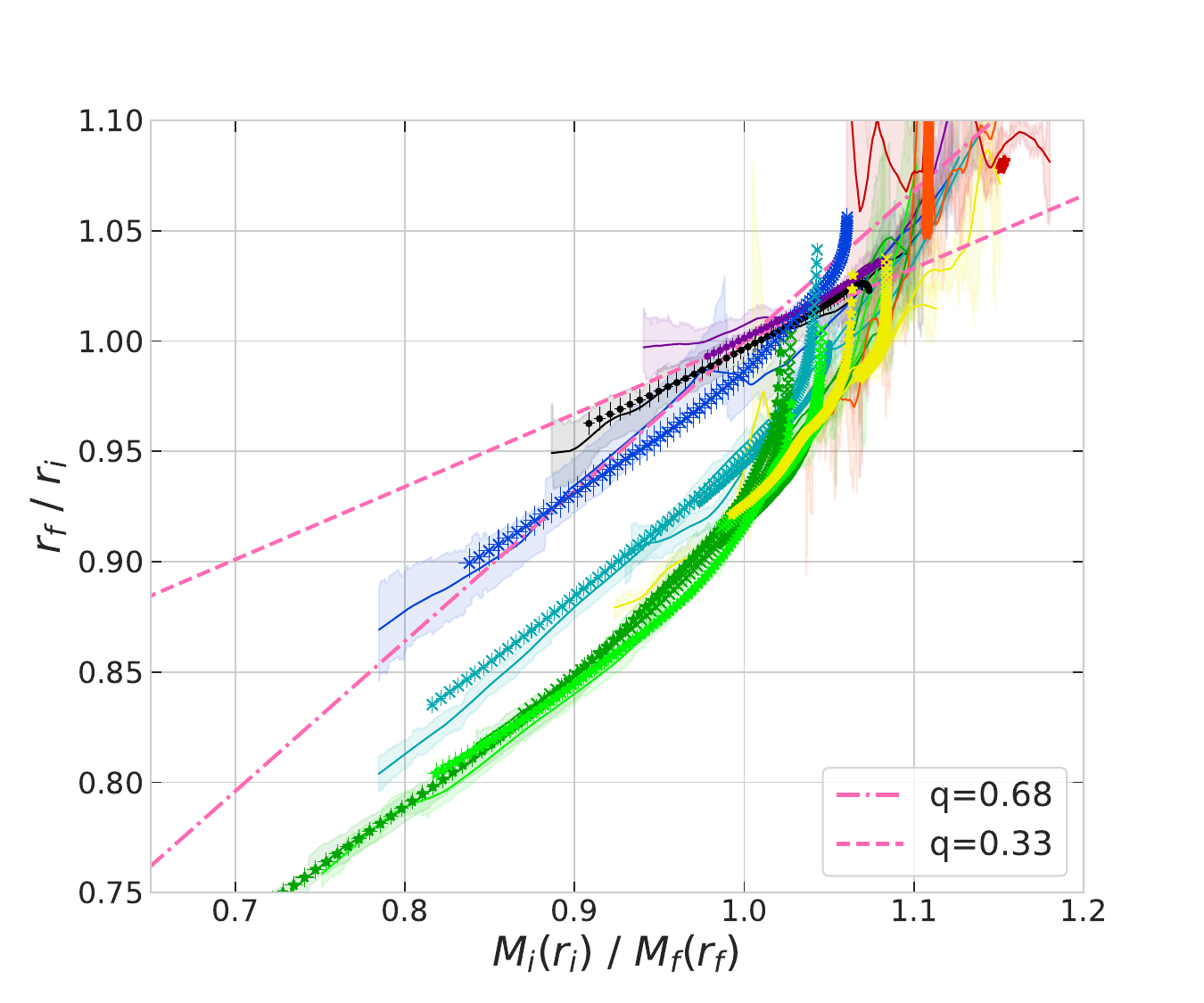}
\begin{minipage}[t]{0.48\linewidth}
    \includegraphics[width=\linewidth]{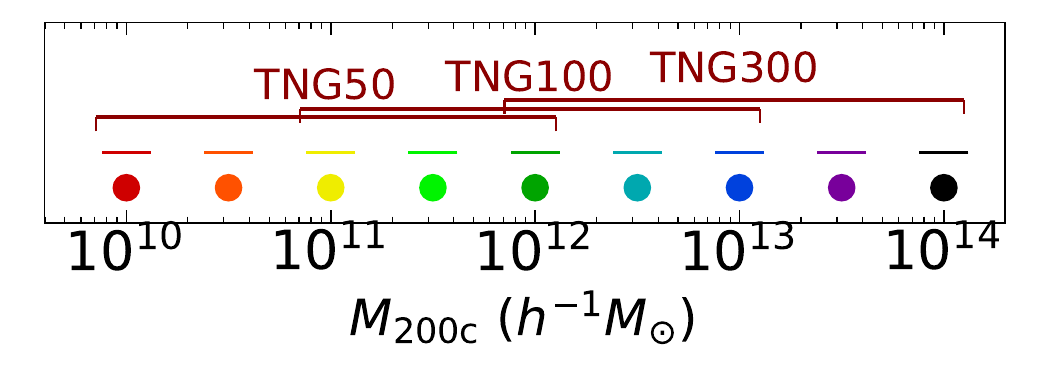}
\end{minipage}%
\begin{minipage}[t]{0.48\linewidth}
    \vspace{-2.2cm}
    \raggedright 
    \red{In all cases, the smaller boxes produce relaxation relations extending to much smaller $M_i/M_f$.}
\end{minipage}
\includegraphics[width=0.48\linewidth,trim={0.5cm 0.5cm 0 1.5cm},clip]{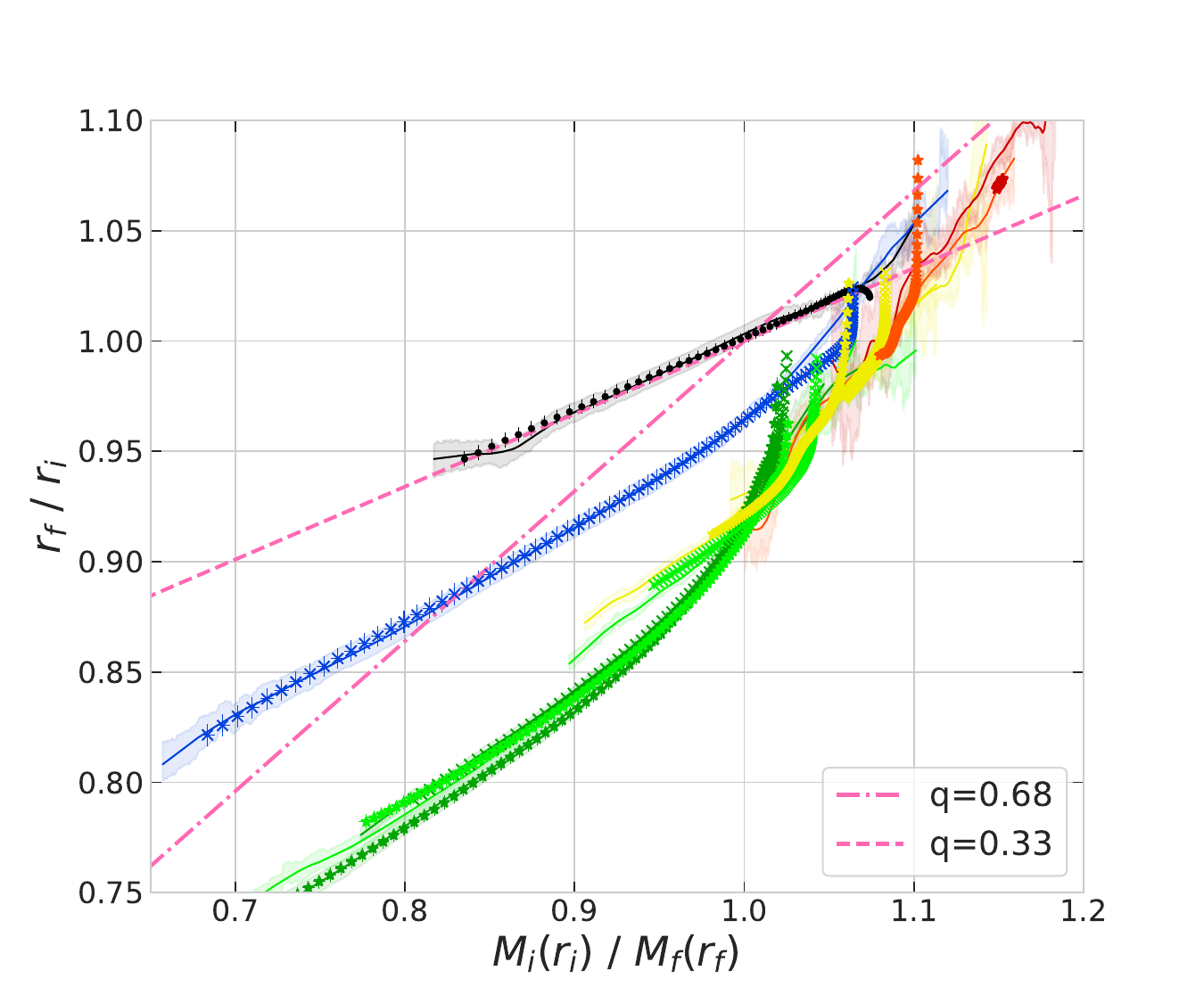}
\includegraphics[width=0.48\linewidth,trim={0.5cm 0.5cm 0 1.5cm},clip]{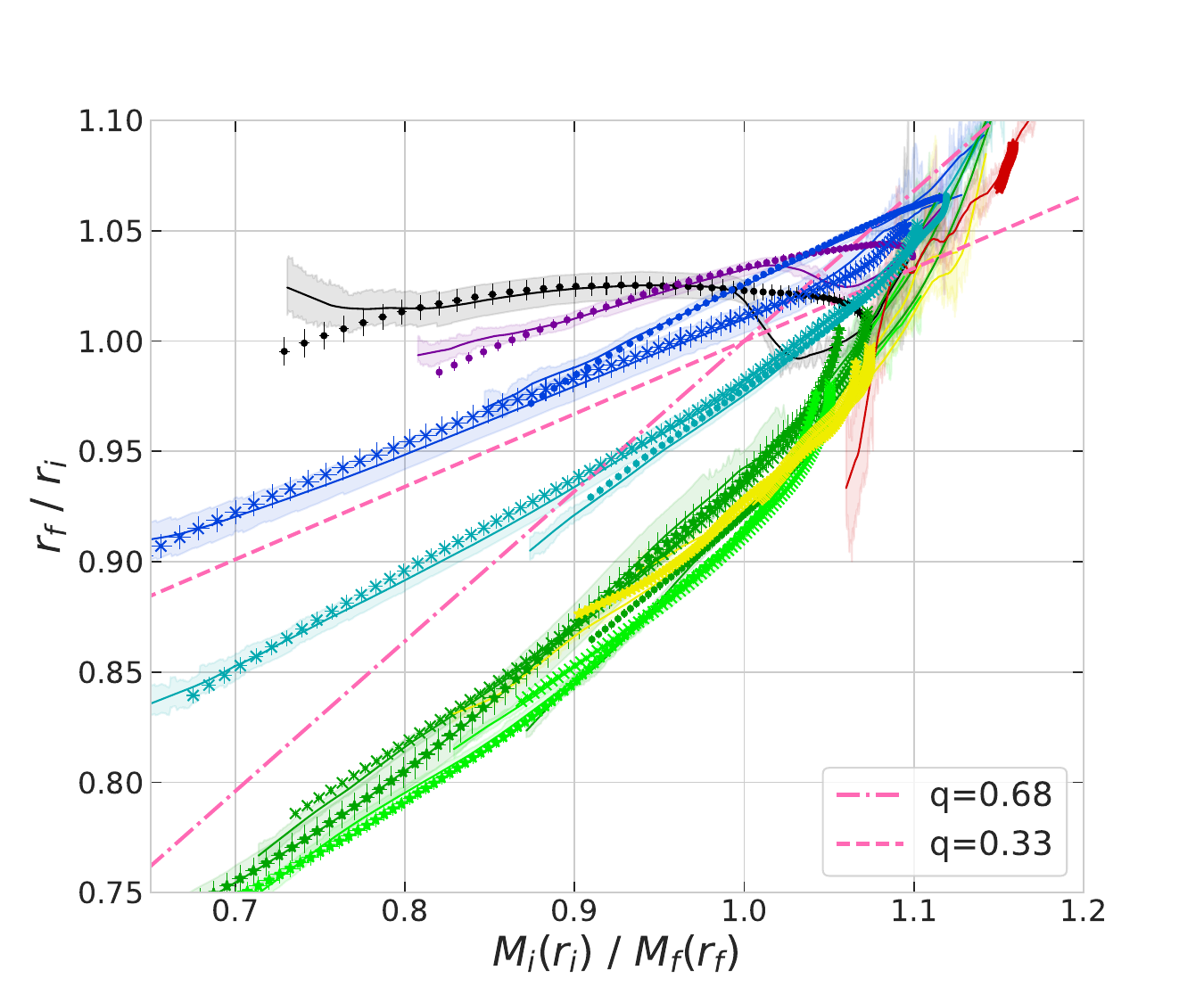}

\caption{The stacked relation between relaxation ratio and mass ratio as a function of halo mass in IllustrisTNG at $z=0$ \emph{(bottom right panel)} and $z=1$ \emph{(other panels)}. In the top right panel, relaxation is shown at $z=1$ for the progenitors of the haloes selected at $z=0$. In the second row, left panel, relaxation is shown at different mass bins at $z=1$, indicated by corresponding mass bins at $z=0$. Points represent stacks over fixed halo-centric distances, and solid lines represent stacks over fixed mass ratios. \oldred{The colour-coding for the halo mass bin is indicated in the inset panel.} The quasi-adiabatic relaxation model \eqref{eq:chi-linear} with $q=0.68$ and $q=0.33$ are shown by the dot-dashed and dashed pink lines, respectively, in each panel. \red{Also see \figref{fig:fit-view-mass-indep-zoomed} for the top left and bottom right plots zoomed in to show the regions close to $M_i/M_f$ and $r_f/r_i$ of unity.}}
\label{fig:fit-view-mass-indep}
\end{figure*}

\begin{figure}
\centering
\includegraphics[width=0.48\linewidth]{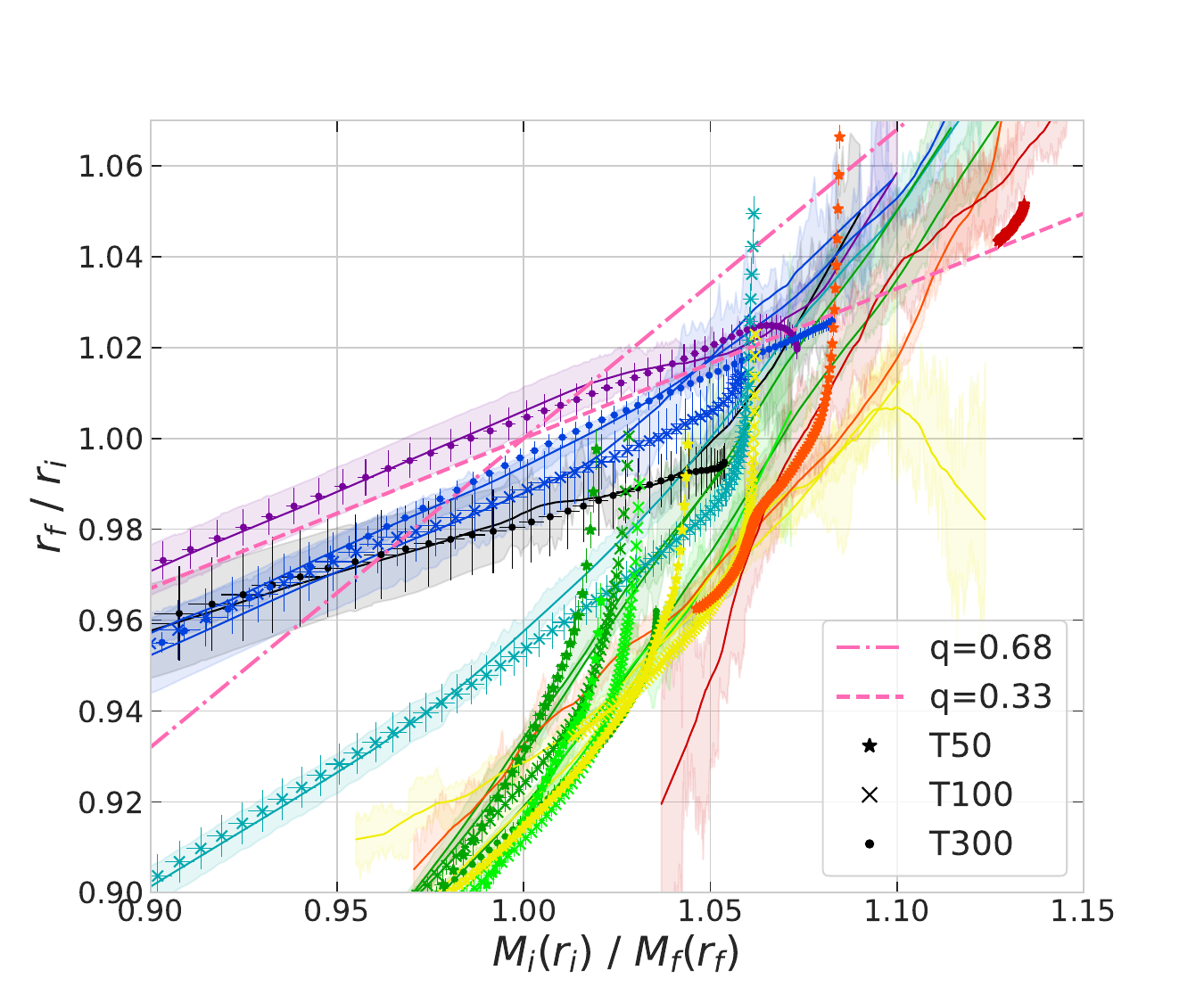}
\includegraphics[width=0.48\linewidth]{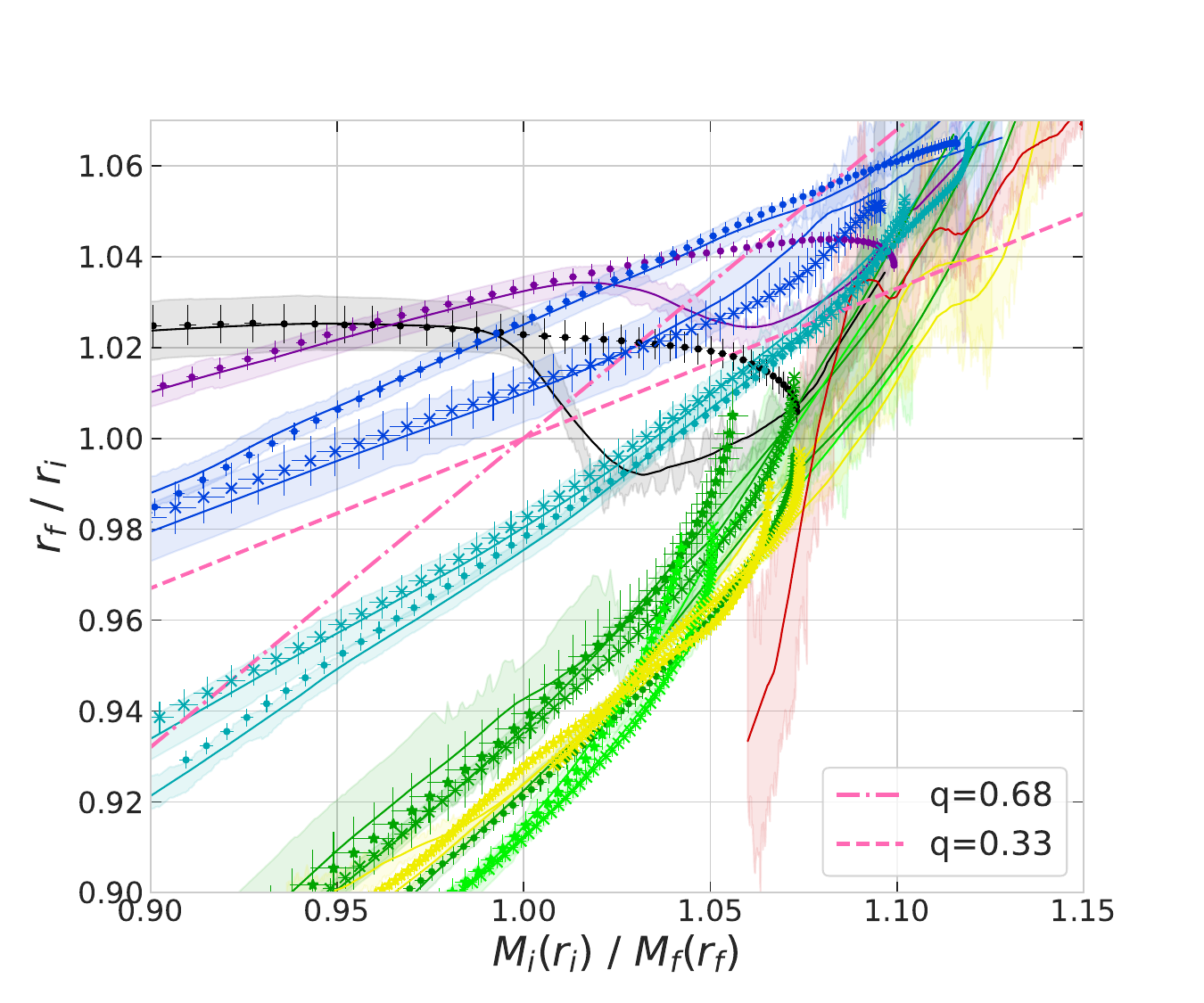}
\caption{\red{Same as \figref{fig:fit-view-mass-indep} but zoomed in to show the regions close to $M_i/M_f$ and $r_f/r_i$ of unity. The left panel shows the zoomed in version of the top left panel of \figref{fig:fit-view-mass-indep} and right panel shows the zoomed in version of the bottom right panel of \figref{fig:fit-view-mass-indep}.}}
\label{fig:fit-view-mass-indep-zoomed}
\end{figure}

For each of the four sets of halo samples, the average relaxation relations ($r_f/r_i$ vs $M_i/M_f$) are shown in \figref{fig:fit-view-mass-indep}, with color indicating only halo mass. \red{We indicate the range of halo masses in each of the three boxes in the legend box.} For a given halo mass, the longest curve extending to small values of $M_i/M_f$ is provided by the smallest box resolving the innermost regions. Notice that the different curves of the same color usually agree with each other, indicating the difference in relaxation relation between different halo masses is larger than those between different simulation boxes for the same halo mass. \red{These differences in the relaxation relation are not isolated to the innermost regions and in some cases the difference is relatively stronger in the outer regions. The outer regions where the change in enclosed total mass and the relaxation effects are minimal is shown in \figref{fig:fit-view-mass-indep-zoomed}. Notice that in some cases the different curves of same color show stronger difference at larger $M_i/M_f$ values. This is particularly true for the $10^{12} \Mh$ haloes between all three TNG boxes, and for $10^{13} \Mh$ haloes between TNG100 and TNG300 boxes. This is likely because the astrophysical modelling in these simulations is resolution dependent and hence leads to different stellar mass fractions and different enclosed total mass profiles $M_f(r_f)$.}

We find that, for a given halo mass, relaxation is usually stronger at the earlier epoch (top left panel) compared to the present time $z=0$ (bottom right panel). \oldred{This can be noticed by comparing the individual curves of the same colour with the model predictions; for example, the dark blue line corresponding to $10^{13}\Mh$ is above the dashed pink line at $z=0$ but below the same line at $z=1$.The amount of relaxation quantified by its deviation from unity, $1-r_f/r_i$ is almost two times larger at $z=1$ than $z=0$.} In both cases, the trend in relaxation with halo mass is similar, with the strongest relaxation observed in $10^{12} \Mh$ haloes. Interestingly, cluster-scale haloes with masses of $10^{14} \Mh$ at redshift $z=1$ exhibit significant relaxation \oldred{see top left panel}, unlike clusters of similar size at the present time.

The progenitors at redshift $z=1$ of the same haloes found at $z=0$ show even stronger relaxation, especially in Milky Way-scale and larger haloes, as shown in the top right panel of \figref{fig:fit-view-mass-indep}. The relaxation follows the simple quasi-adiabatic model \eqref{eq:chi-linear} with $q=0.33$ among larger cluster-scale ($10^{14} \Mh$) haloes, while group-scale haloes are consistent with the second-order polynomial relation proposed by Abadi et al. (2010) \cite{2010MNRAS.407..435A}. In the bottom left panel, the relaxation relation is shown for all haloes within a narrow mass bin around the median mass of the progenitors of the haloes selected at $z=0$. Notice that the relaxation relation shifts further lower in Milky Way-scale haloes and smaller clusters with the inclusion of those additional haloes.

\begin{figure}[htbp]
\centering
\includegraphics[width=0.48\linewidth,trim={0.5cm 0 0 0},clip]{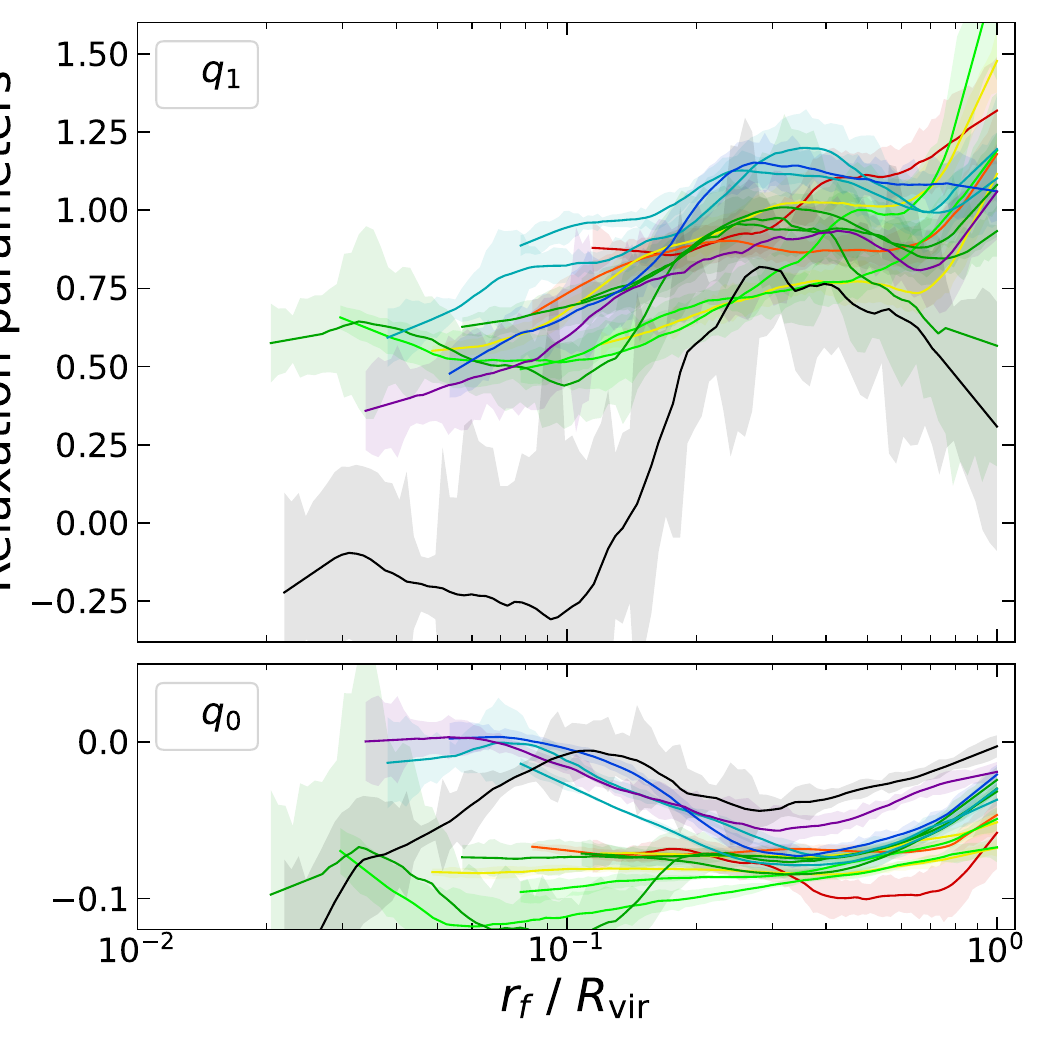}
\includegraphics[width=0.48\linewidth,trim={0.5cm 0 0 0},clip]{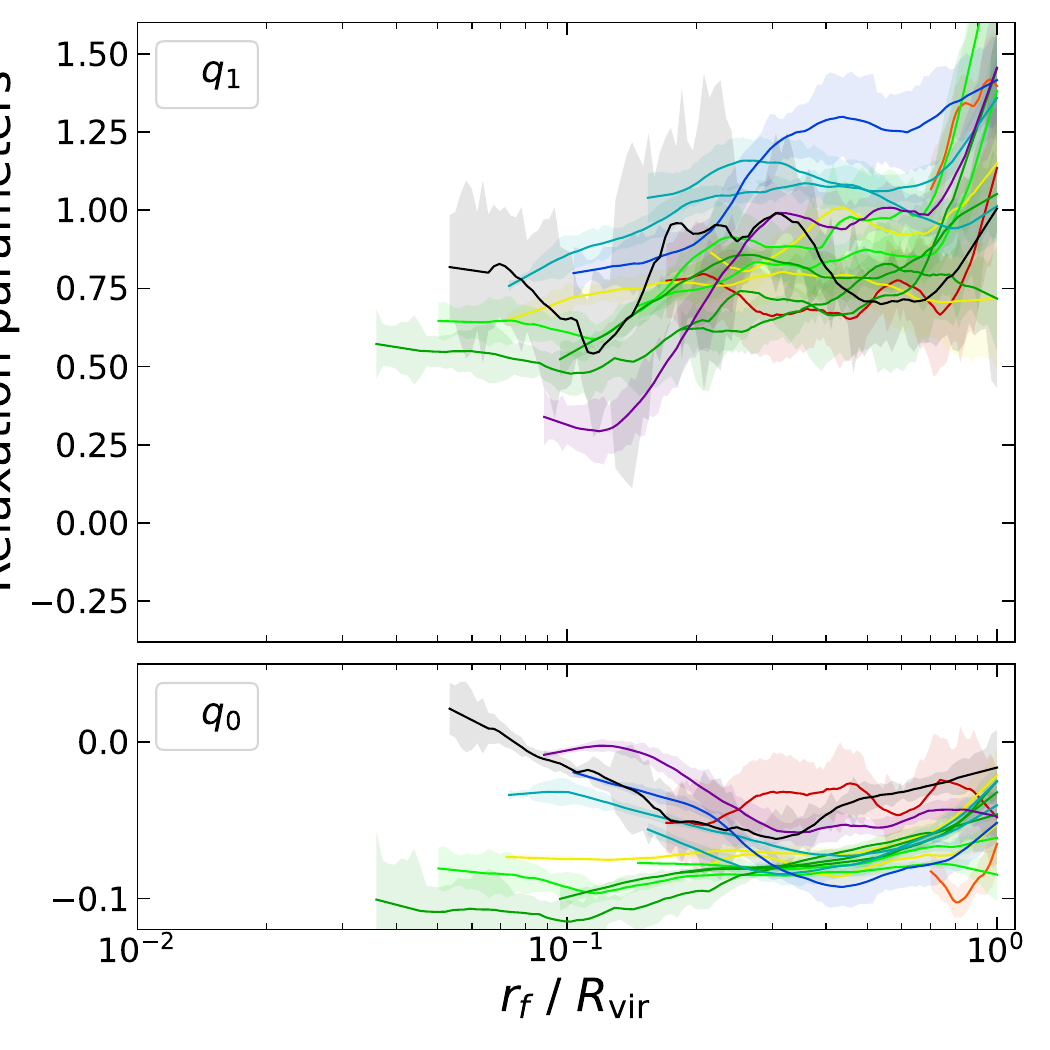}
\includegraphics[width=0.48\linewidth,trim={0.5cm 0 0 0},clip]{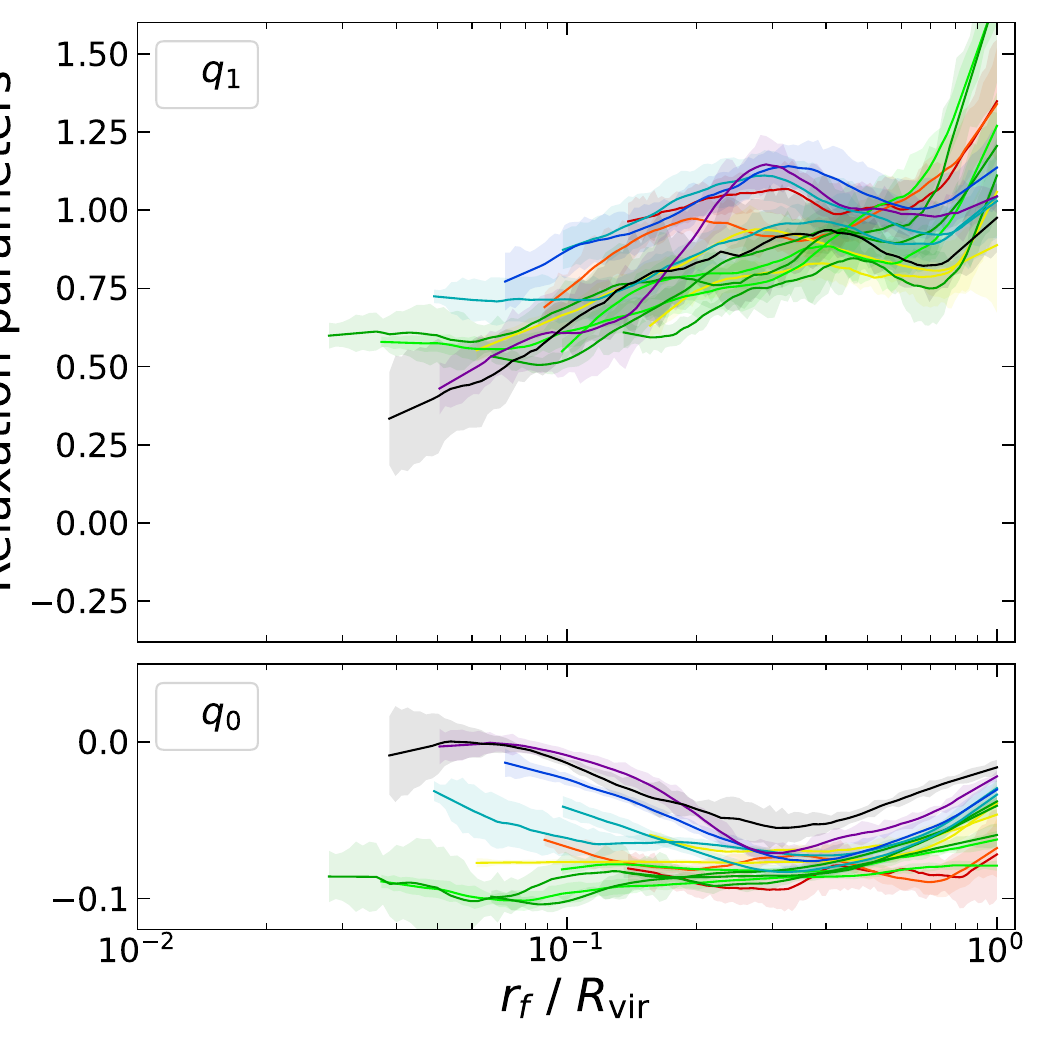}
\includegraphics[width=0.48\linewidth,trim={0.5cm 0 0 0},clip]{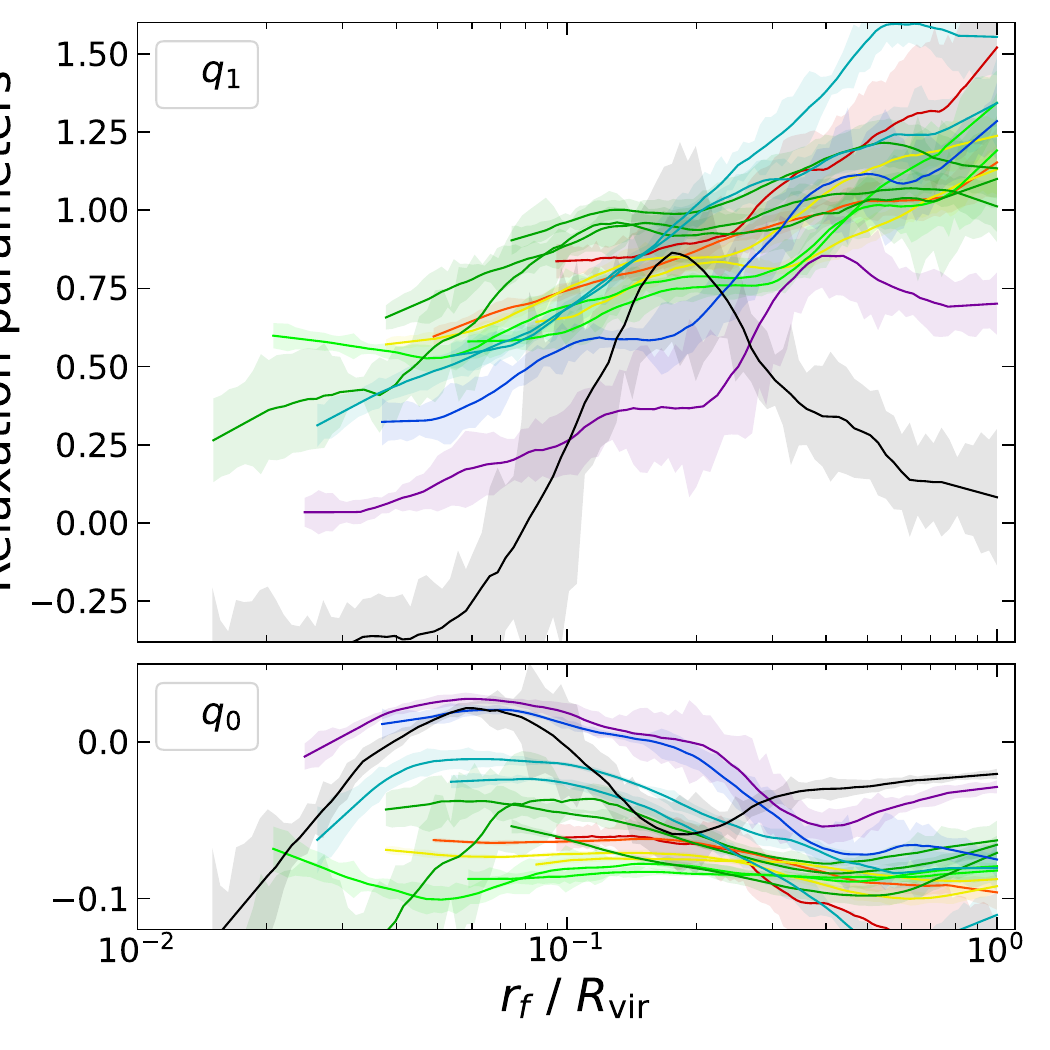}
\makebox[0pt][r]{%
    \raisebox{8em}{%
      \includegraphics[width=.28\linewidth]{plots/Mass_bin_labels_enlarged.pdf}%
    }\hspace*{21em}%
  }%
\caption{Linear quasi-adiabatic relaxation model parameters $q_1$ and $q_0$ as a function of the halo-centric distance at different halo masses in IllustrisTNG at $z=0$ \emph{(bottom right panel)} and $z=1$ \emph{(other panels)}. In the top right panel, relaxation is shown at $z=1$ for the progenitors of the haloes selected at $z=0$. In the second row, left panel, relaxation is shown at different mass bins at $z=1$, indicated by corresponding mass bins at $z=0$. \oldred{The colour-coding for the halo mass bin is indicated in the inset panel in the lower left panel.}}
\label{fig:rf-fit-params}
\end{figure}

\subsubsection*{Halo-centric dependent relaxation}

We have tested the locally linear model of relaxation relation \eqref{eq:chi-linear-q0} with our halo samples at $z=1$ and found it to hold reasonably well. For each halo sample, at each halo-centric distance $r_f$, the relationship between mass ratio and relaxation ratio across all haloes is best fitted by a linear curve with the slope and offset parameters namely $q_1(r_f)$ and $q_0(r_f)$. The radial profiles of these two relaxation parameters are shown for the four set of halo populations in \figref{fig:rf-fit-params} with the color-coding given by \figref{fig:mass_bin_label-z01} for the associated halo mass.

Notice that the universality in these relaxation profiles extends to much larger mass haloes ($\sim 10^{13.5} \Mh$) at $z=1$ (top left panel) compared to $z=0$ (bottom right panel). This is despite the fact that such massive haloes are even rarer at $z=1$ than at $z=0$, as indicated by the value of $\nu$ in \figref{fig:mass_bin_label-z01}. For all the halo samples selected at $z=0$, their traced progenitor populations at $z=1$ show nearly universal relaxation profiles, with the exception of $q_0(r_f)$ in the inner halo (shown in the top right panel). This universality is even more apparent among the halo populations selected in narrow mass bins at $z=1$ based on the median masses of the progenitor populations (see bottom left panel).

The less universal and relatively noisier profiles in direct progenitors might be due to the inclusion of low mass fast accreting haloes together with high mass slow accreting haloes or simply a systematic issue such as a few misidentified progenitors. This difference between the direct progenitor populations and the haloes selected by the median progenitor masses seems to be more noticeable in the relaxation profiles than in the radially independent relaxation relations. For example, the black curves (corresponding to $10^{14} \Mh$) are noticeably different especially in inner regions between top right and bottom left panels in \figref{fig:rf-fit-params}, however, the overall relaxation relation shown in \figref{fig:fit-view-mass-indep} is well-fitted by the linear $q=0.33$ model in both the panels. %
While the exact reason is not yet clear, we note that the radially-dependent relaxation might be more sensitive to this difference. 
Another interesting thing to note is that $q_1(r_f)$ shows a small non-monotonic behavior in the outer haloes at all masses in $z=1$ (e.g., see lower left panel) that is not seen at $z=0$ (lower right panel).
The range of halo-centric distances, over which $q_1(r_f)$ is linearly increasing with $\log \left( \frac{r_f}{R_{\rm{vir}}} \right) $, referred to as the loglinear regime seem to increase with time. Investigations of dynamics of the response through time-series analysis (presented in a separate publication \cite{2024arXiv240708030V}) suggest that these are likely due to the astrophysical effects that are seen first in the inner halo and then in the outer halo. %

\begin{figure}[htbp]
\centering
\includegraphics[width=0.6\linewidth]{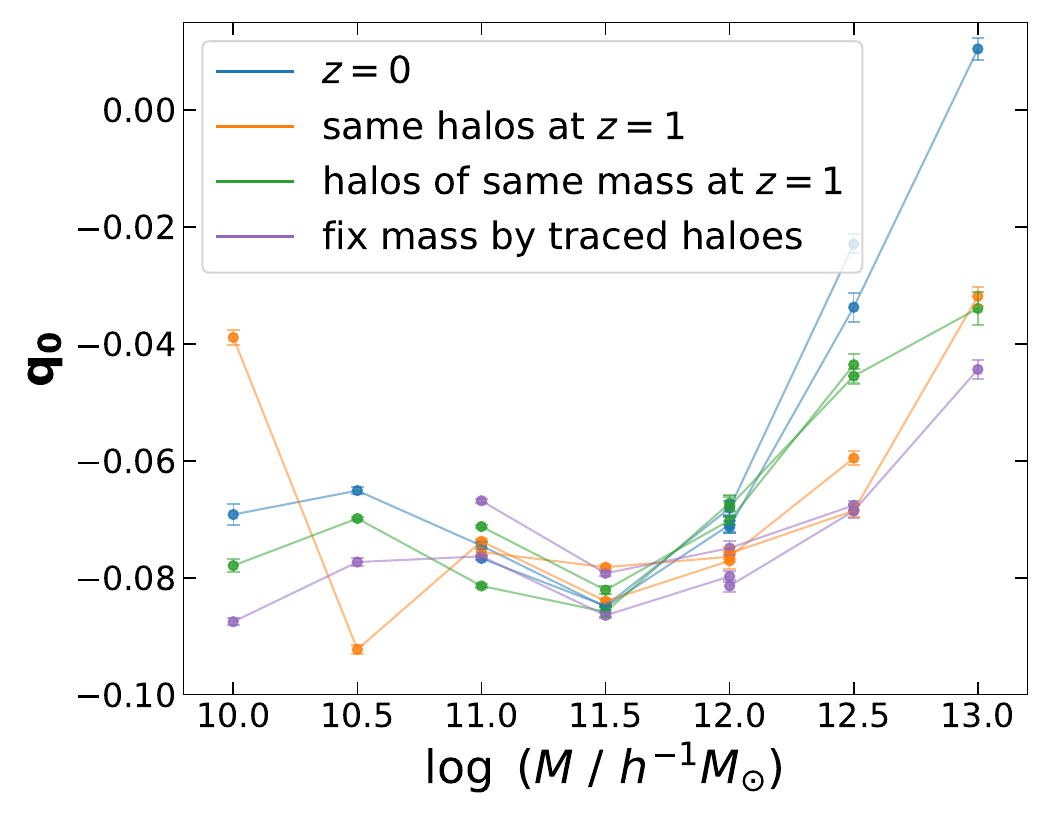}
\caption{Mean of the radially dependent quasi-adiabatic relaxation offset, $q_{0}$ as a function halo mass in the four sets of halo samples indicated by color.}
\label{fig:fit-fit-func-q}
\end{figure}

The offset parameter $q_0$ shown in the lower sub-panels is relatively uniform across the halo, especially among the low-mass haloes. The \figref{fig:fit-fit-func-q} shows the mean of $q_0$ in all four sets of halo samples. The $q_0$ parameter is usually more negative across all $z=1$ halo populations compared to $z=0$, indicating a stronger relaxation offset. Additionally, the values are more universal with halo mass at $z=1$. A negative value of $q_0$ is expected to be a result of recent feedback outflows \cite{2023Velmani&Paranjape}. These feedback outflows are produced by a combination of AGN and stellar feedbacks. In the high mass haloes, powerful AGN feedbacks at earlier epochs lead to significant suppression in the star formation activity reducing the stellar feedbacks at present epoch. The reduction in the magnitude of $q_0$ from $z=1$ to $z=0$ in the high mass haloes could be simply due to the reduction in overall feedback among those haloes.
~%

\subsection{Role of astrophysical modelling on the relaxation response}
In this section, we investigate the role of various feedback parameters prescribed in the IllustrisTNG simulations using the set of CAMELS simulations performed with the IllustrisTNG model. \oldred{And we also investigate the physics variation simulations from the EAGLE suite described in \secref{sec:sims-EAGLE} on the role of different components of the astrophysical modelling such as the nature of the gas and star formation mechanisms in addition to the feedback implementations. We first describe and present our results and then discuss all these results below.}

\paragraph{Dependence on feedback strengths using CAMELS simulations}
\label{sec:res-physvar-CAMELS}
This includes a set of 41 hydrodynamical simulations, with one replicating the reference TNG model in a smaller cosmological volume, and 10 simulations each by varying 4 different feedback parameters as described in \secref{sec:sims-CAMELS}. To recall, this includes two supernovae feedback parameters ($A_{\mathrm{SN1}}$ and $A_{\mathrm{SN2}}$) and another two AGN feedback parameters ($A_{\mathrm{AGN1}}$ and $A_{\mathrm{AGN2}}$).
Due to the limited resolution and the smaller volumes of the CAMELS simulation, among all the mass bins shown in \figref{fig:mass_bin_label-z01}, we consider only $10^{11} \Mh$, $10^{11.5}$, and $10^{12} \Mh$ at both redshifts $z=0$ and $z=1$. Even in these mass bins, we consider only the outer well-resolved regions of the haloes. Also, since the cosmological volume is smaller than even the smallest TNG50 simulation, we have only a smaller sample of haloes at each of these halo mass bins. We find this sample size insufficient to estimate the radially-dependent relaxation parameters at each of these mass bins. We consider the following two approaches to alleviate this issue.

\subsubsection*{(i) Intercepts in the Relaxation Relation}

The intercepts of the relaxation relation, given by the relation between $M_i/M_f-1$ and $r_f/r_i-1$, already provide interesting information about the relaxation. For example, the y-intercept denotes the offset in the relaxation ratio $r_f/r_i$ from unity for the shells having a mass ratio of unity $M_i/M_f=1$. Similarly, the x-intercept denotes the offset in the mass ratio $M_i/M_f$ from unity for the shells having a relaxation ratio $r_f/r_i = 1$.

In a given sample of haloes, we denote the average x and y intercepts as $q_x$ and $q_y$ respectively. For example, if we consider the Milky Way scale haloes at redshift $z=0$, the relaxation relation shown by the green curves in the lower right panel of \figref{fig:fit-view-mass-indep} indicates that $q_x$ will be positive and $q_y$ will be negative. In general, we expect that feedback effects would lead to larger $q_x$ and more negative $q_y$.

\begin{figure}[htbp]
\centering
\includegraphics[width=\linewidth]{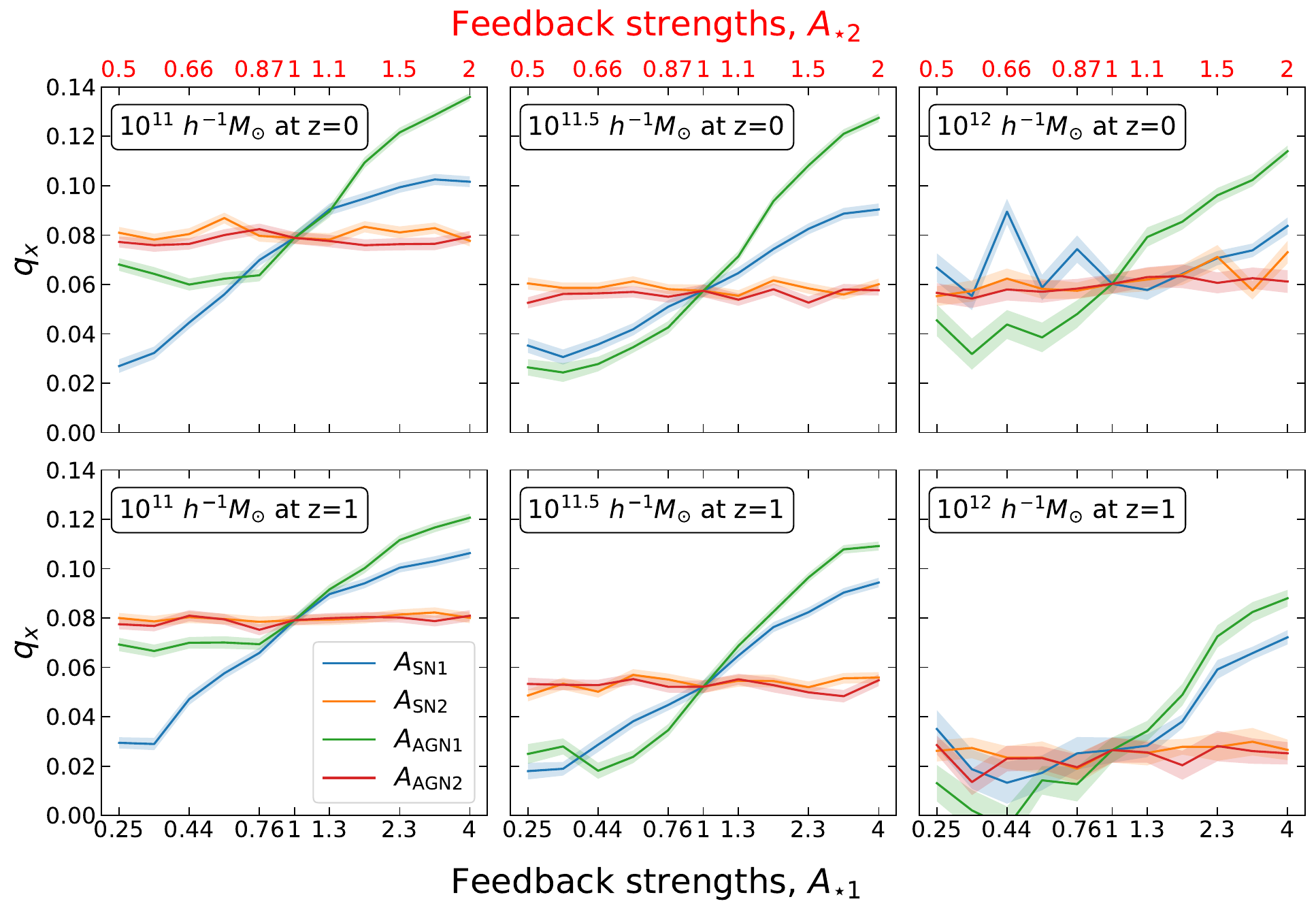}
\caption{Relaxation offset parameter $q_x$ as a function of the baryonic astrophysical feedback parameters in haloes found in CAMELS-TNG at three different halo masses. Top: $z=1$, Bottom: $z=0$}
\label{fig:camels-qx0}
\end{figure}

Due to the limited resolution of the CAMELS simulation, the outer well-resolved regions in most haloes didn't have a mass ratio less than unity. This makes the y-intercept of the relaxation relation available only in a smaller number of haloes. This makes our estimation of $q_y$ very noisy, and hence we only interpret the parameter $q_x$ in these simulations. This parameter $q_x$ is presented in \figref{fig:camels-qx0} as a function of the astrophysical parameters in the CAMELS TNG set of simulations at both redshifts $z=0$ and $z=1$.

\subsubsection*{(ii) Wider mass bin}

\begin{figure}[htbp]
\centering
\includegraphics[width=.9\linewidth]{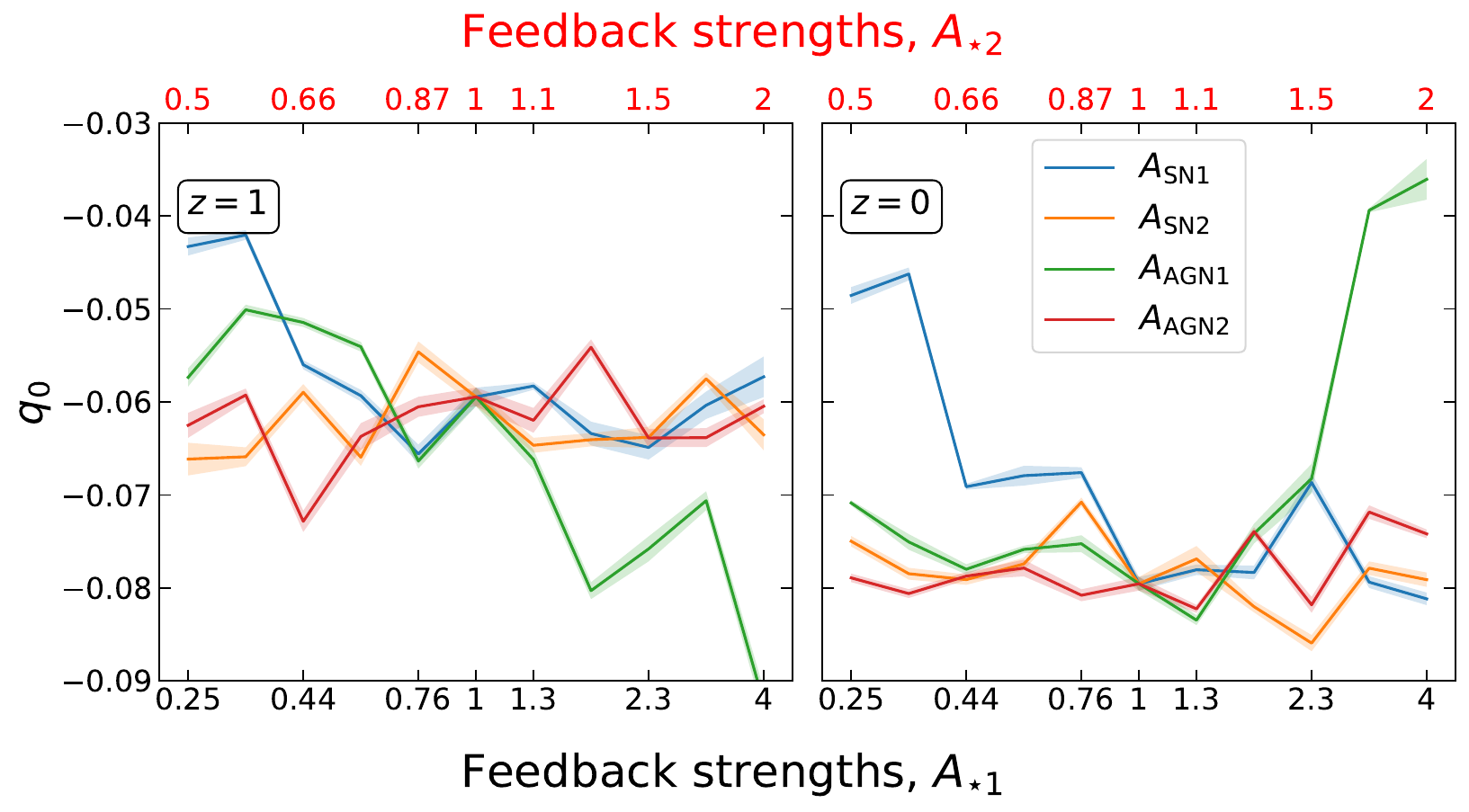}
\caption{Relaxation offset parameter $q_0$ as a function of the baryonic feedback parameters in CAMELS-TNG. Left: $z=1$, Right: $z=0$.}
\label{fig:camels-q0q1}
\end{figure}

We found that the radially-dependent relaxation parameters are usually more uniform across a wider range of halo masses \oldred{in the original IllustrisTNG simulation}. We leverage this to consider haloes in a wider mass bin from $10^{11} \Mh$ to $10^{12} \Mh$, which gives a sufficient number of haloes to obtain the radially-dependent relaxation model parameters. However, still, the radial range is not sufficient to accurately model the radial dependence of the slope parameter $q_1(r_f)$, so we only investigate the offset parameter $q_0$, defined as the mean of the $q_0(r_f)$. This parameter $q_0$ is usually negative, and it is expected to be more negative when the offset produced by overall feedback effects is stronger. This model-dependent offset parameter is presented in \figref{fig:camels-q0q1} at both redshifts $z=0$ and $z=1$.

\begin{figure}[htbp]
    \centering
    \includegraphics[width=0.32\linewidth]{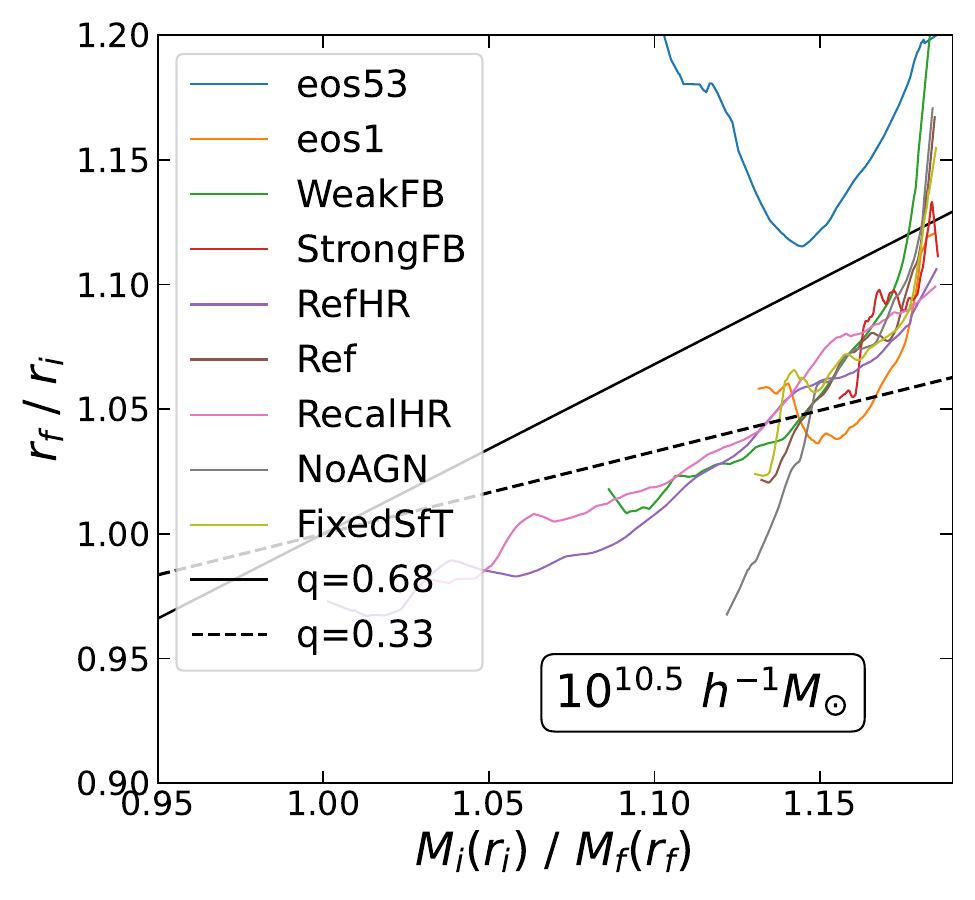}
    \includegraphics[width=0.32\linewidth]{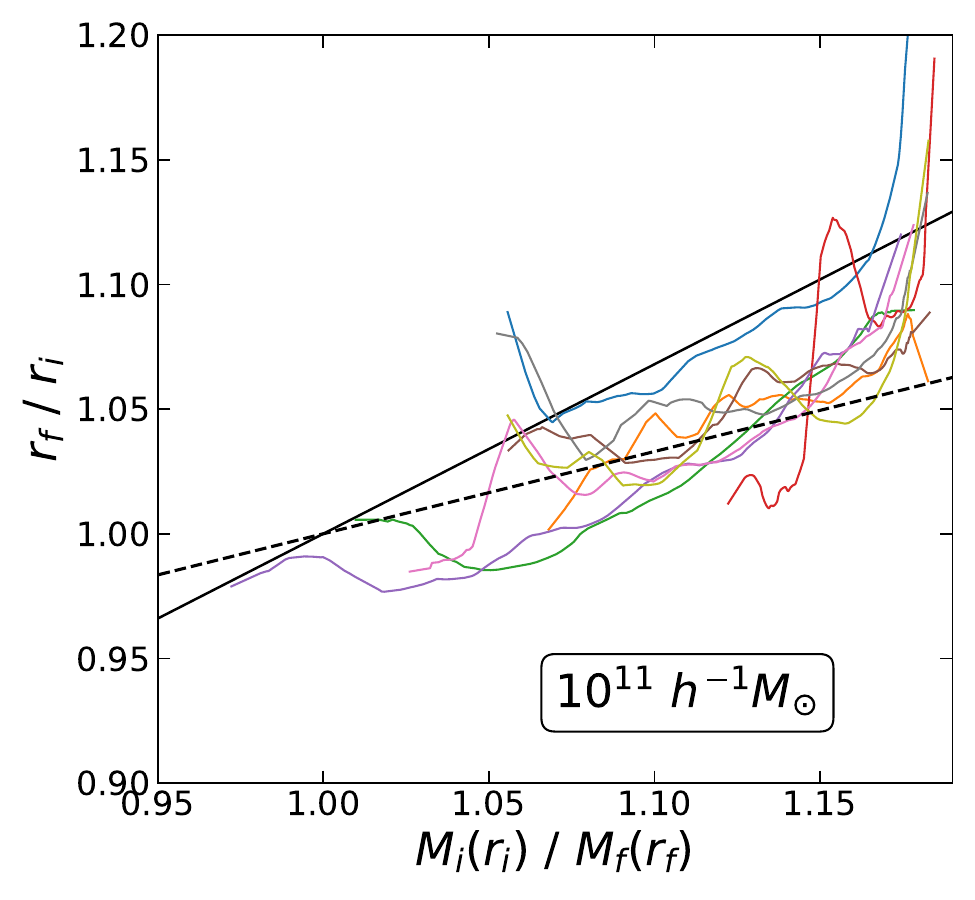}
    \includegraphics[width=0.32\linewidth]{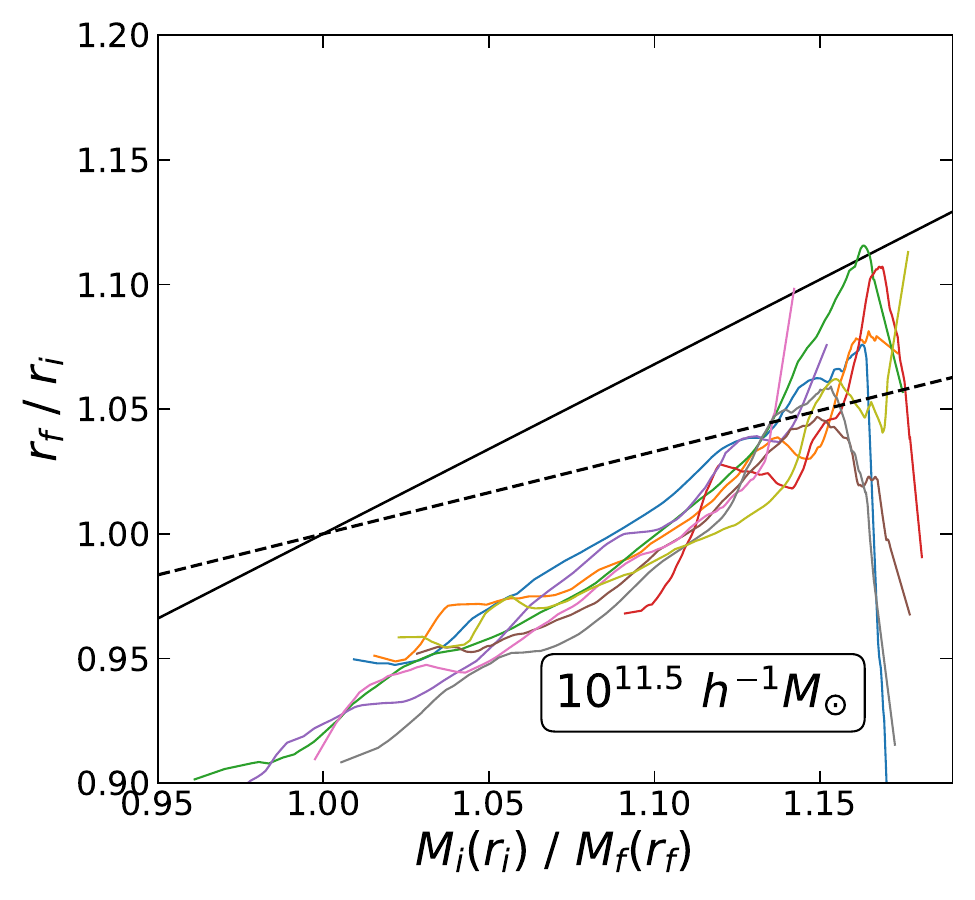}
    \caption{Relaxation relation in the physics variation EAGLE simulations for haloes in the mass bins from $10^{10.5} \Mh$ to $10^{11.5} \Mh$. Here colors represent the specific simulation with a variation in the baryonic physics prescription.}
    \label{fig:EAGLE-rad-indep}
\end{figure}

\begin{figure}[htbp]
    \centering
    \includegraphics[width=0.7\linewidth]{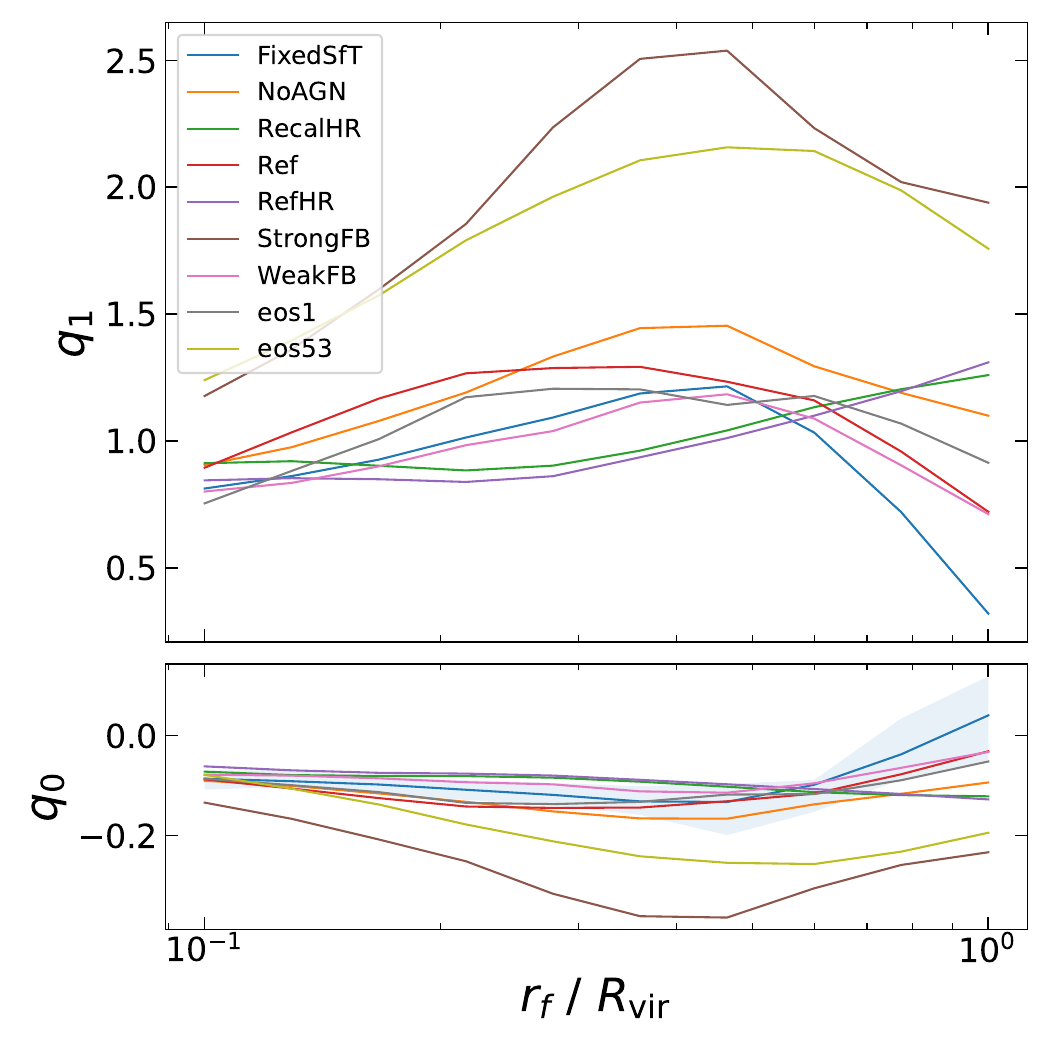}
    \caption{Radially-dependent relaxation parameters for low-mass haloes from $10^{10.5} \Mh$ to $10^{11.5} \Mh$ as a function of the halo-centric distance in the physics variation EAGLE simulations. Here colors represent the specific simulation with a variation in the baryonic physics prescription.}
    \label{fig:EAGLE-rad-dep}
\end{figure}

\paragraph{Effect of astrophysical model variations in EAGLE simulations}
These physics variation simulations from the EAGLE suite are described in \secref{sec:sims-EAGLE}. \oldred{Since these simulations have nearly ten times higher mass resolution than CAMELS, we also study dwarf scale haloes of masses including $10^{10.5} \Mh$. However, since the box size is more than three times smaller than the CAMELS, we don't study the $10^{12} \Mh$ bin with very few haloes.} We consider haloes in three different mass bins centered at $10^{10.5} \Mh$, $10^{11} \Mh$, and $10^{11.5} \Mh$. In these narrow mass bins, the mean relaxation relation obtained by stacking independent of the halo-centric distance is shown in \figref{fig:EAGLE-rad-indep} for these physics variation simulations. In \figref{fig:EAGLE-rad-dep}, we present the radially dependent relaxation parameters for a larger sample of haloes in a wider range of halo masses from $10^{10.5} \Mh$ to $10^{11.5} \Mh$.

\subsubsection{Discussion}
\paragraph{Effect of feedback strength:}
Let us first focus on the dependence of relaxation response on the feedback parameters in CAMELS. Among all haloes investigated, \oldred{note that in \figref{fig:camels-qx0},} the feedback strength parameters $A_{\mathrm{SN1}}$ and $A_{\mathrm{AGN1}}$ have a strong influence on the relaxation, with $q_x$ typically increasing monotonically when increasing these parameters. On the other hand, 
the wind speed parameters $A_{\mathrm{SN2}}$ and $A_{\mathrm{AGN2}}$ have negligible effect on the relaxation characterized by both $q_x$ and $q_0$.

Recall that when varying only a wind speed parameter, it affects the burstiness of the feedback outflows while keeping the overall \oldred{energy} flux constant. Suppose the deviations in the relaxation relation from the idealized adiabatic model, quantified by non-zero offset parameters, are caused by the transfer of angular momentum between the dark matter particles and the baryonic particles. In that case, one may expect that the nature of the baryonic feedback quantified by the wind speed parameters will have a significant influence on the value of $q_x$. However, our results suggest otherwise.

In Velmani \& Paranjape (2023) \cite{2023Velmani&Paranjape}, we have argued that the relaxation offset is a reflection of the fact that the dark matter shells have not yet expanded in response to the recent feedback gas outflows. These results suggest that the time taken for the dark matter shells to expand are sufficiently long enough, that only the overall gas outflow is relevant irrespective of the speed and burstiness. 

\oldred{Notice that in \figref{fig:camels-qx0}} the AGN feedback strength generally has a stronger influence on the relaxation among the high-mass haloes, whereas the supernova feedback strength has a stronger influence in lower-mass haloes. This is consistent with our expectations that AGN feedbacks dominate in more massive haloes. However, at all masses, AGN feedback starts dominating the value of $q_x$ when their strengths are set to be higher than the reference model of TNG; this is indicated by the green curves in \figref{fig:camels-qx0}. This suggests that the relative importance of AGN and supernovae feedback on the dark matter relaxation depends strongly on their implementation.

While $q_x$ which characterizes the relaxation offset predominantly in the outer halo, is always larger with the stronger implementation of AGN feedback at both \( z=0 \) and \( z=1 \), the value of $q_0$, which is averaged over both inner and outer halo, shows a different trend.  Notice, in \figref{fig:camels-q0q1}, that the stronger AGN feedback implementation leads to a weaker relaxation offset at \( z=0 \) and a stronger offset at \( z=1 \). \oldred{To understand this, we have to consider the inter-connected role of SN and AGN feedback parameters as shown by Tillman et al. (2023) \citep{2023TillmanBurkhart_etal}. Stronger AGN feedbacks in the past would suppress the star formation, causing a} overall reduction in total feedback at \( z=0 \). This could have led to reduction in the relaxation offset strength in the inner haloes, the outer halo has not yet responded by $z=0$. These results highlight the significance of feedback mechanisms in building a physical understanding of dark matter halo relaxation.

\paragraph{Effect of variation in overall astrophysical model:}

Let us now delve into the relaxation results from small boxes of EAGLE simulations on the effect of astrophysical model variation.
We find that the deviation in the relaxation across different astrophysical modelling is usually smaller than the differences from one halo mass to the other \oldred{with few exceptions as shown in \figref{fig:EAGLE-rad-indep}}. Notice that the gas equation of state has a strong influence on the relaxation relation, especially among low-mass haloes. In particular, stiffer equations of state lead to a very large $r_f/r_i$ indicating a stronger expansion of the dark matter shells in response to galaxy formation. Similar trends have been found in simpler self-similar systems of individual haloes \cite{2024VelmaniParanjape}. 

\oldred{The NoAGN run shows some interesting features in the relaxation. At $10^{10.5}\Mh$, the grey curve in the left panel of \figref{fig:EAGLE-rad-indep} shows the steepest decrease in $r_f/r_i$ with a decrease in $M_i/M_f$, indicating a strong contraction at such high mass ratio of $M_i/M_f \sim 1.12$. At $10^{11.5}\Mh$, the grey curve is clearly lower than all other curves for most of the mass ratios, once again suggesting a stronger relaxation in the total absence of AGN feedback.}

\oldred{With stronger supernova feedback implementation, the mass ratios corresponding to the same halo-centric distances explored tend to be much larger than all other runs. However, interestingly the relation between relaxation ratio $r_f/r_i$ and mass ratio $M_i/M_f$ seems to follow the same curve as other runs, albeit more noisy in some cases.}

\oldred{Switching our focus to the } radially dependent relaxation parameters \oldred{shown in \figref{fig:EAGLE-rad-dep}}. These results also reflect the strong influence of the equation of state of gas on the relaxation response of dark matter haloes. Additionally, these results highlight the effect of supernovae feedback modeling. In particular, the $q_0(r_f)$ is more negative, indicating a stronger offset in the haloes found in the simulation with stronger supernova feedback implementation (brown curve vs. rose curve).

\section{Conclusion}
\label{sec:conclusion}

In this work, we investigated the influence of astrophysical modeling on the relaxation response of dark matter haloes at different epochs, specifically focusing on \( z=0 \) and \( z=1 \). The analysis is divided into three main parts, each shedding light on the role of various astrophysical processes in shaping the dark matter content of haloes.

We began by examining the relaxation response at an earlier redshift (\( z=1 \)) in the IllustrisTNG simulations using three distinct sets of halo samples, which highlight the variations in relaxation across different halo masses. Our study reveals that dark matter relaxation tends to be stronger (smaller \( r_f/r_i \)) at the earlier epoch compared to the present among haloes of the same mass. This is even more prominent among the progenitors of present epoch haloes. Notably, we observe that cluster-scale haloes at \( z=1 \) show significant relaxation (\( r_f/r_i < 1 \)) that is also a function of the change in the enclosed mass (\( M_i/M_f \)). This is in contrast to similar haloes at the present epoch, where \( r_f/r_i \) stayed close to unity on average irrespective of the value of \( M_i/M_f \).

We also find that the locally linear quasi-adiabatic relaxation model is a good description of the relaxation relation at this earlier epoch, demonstrating its robustness in capturing the dark matter response across redshifts. Moreover, the parameters of the radially dependent relaxation are found to be more universal across a much wider range of masses at \( z=1 \). For example, the progenitors of even the most massive clusters are well characterized by the simple three-parameter model of relaxation that was developed with a focus on galactic-scale haloes at \( z=0 \). This suggests that the deviation from the three-parameter model is a result of late-time events such as mergers in the cluster scale haloes.%

Next, we explored variations in astrophysical feedback strengths within the IllustrisTNG model using simulations from the CAMELS project, which varies four different feedback parameters: two for stellar feedback and two for AGN feedback. Our analysis shows that the parameters controlling the energy flux of the feedback have a significant impact on the relaxation of dark matter at different epochs. In contrast, the parameters governing the speed and burstiness of feedback have negligible effects on the halo relaxation response. This further strengthens our argument that the relaxation offset is caused by the dark matter shells that have not yet responded to the recent feedback outflows. %

We find that variations in stellar feedback strengths have a larger impact among dwarf galaxy-scale haloes, while variations in AGN feedback parameters exert a stronger influence on Milky Way-scale haloes. Notably, the relaxation offset in the outer well-resolved regions is stronger at the present epoch than at \( z=1 \), contrasting with results from the inner regions explored in the IllustrisTNG simulations.
The stronger implementation of AGN feedback tends to result in greater relaxation at both \( z=0 \) and \( z=1 \) in the outer regions of the haloes. However, in the slightly inner regions, stronger AGN feedback implementation leads to a weaker relaxation offset at \( z=0 \) and a stronger offset at \( z=1 \). We interpret this as a consequence of the overall reduction in total feedback at \( z=0 \) due to the suppression of star formation caused by higher AGN feedbacks in the past. \oldred{This could also explain the reduction in the relaxation offset in the cluster scale haloes from redshift $z=1$ to $z=0$ as discussed in \secref{sec:res-itng-z01}}. These results highlight the significance of feedback mechanisms in building a physical understanding of dark matter halo relaxation.

Finally, we assessed the impact of different astrophysical models in the EAGLE simulations. Supernova feedback strengths show a similar trend to that observed in the CAMELS simulations. Additionally, we find that the gas equation of state has the strongest effect on the relaxation response of dark matter, which is consistent with expectations from self-similar models of isolated galaxies with their host haloes.

These results are directly applicable in interpreting the upcoming large-volume surveys, in accurately modelling the astrophysical effects on dark matter distribution with a few parameters. In a separate publication \cite{2024arXiv240708030V}, we investigate the dynamical evolution of these effects through time-series analysis of the evolutionary history of a variety of haloes. In a future work, we plan to incorporate the associated time scales found in that analysis, with the results presented here, to build a unified model of the dynamical response of the dark matter within haloes to the galaxies they host.

\section*{Data availability}
No new data were generated during the course of this research.

\acknowledgments
AP's research is supported by the Associates scheme of ICTP, Trieste. We gratefully acknowledge the use of the high-performance computing facilities at IUCAA (\url{http://hpc.iucaa.in}). This work benefited significantly from the use of several open-source software packages, including NumPy \citep{vanderwalt-numpy},\footnote{\url{http://www.numpy.org}} SciPy \citep{scipy},\footnote{\url{http://www.scipy.org}} Matplotlib \citep{hunter07_matplotlib},\footnote{\url{https://matplotlib.org/}} Pandas \citep[][]{reback2020pandas},\footnote{\url{https://pandas.pydata.org/about/}} Schwimmbad \citep{schwimmbad},\footnote{\url{https://joss.theoj.org/papers/10.21105/joss.00357}} H5py,\footnote{\url{https://www.h5py.org/}} Colossus \citep{colossus},\footnote{\url{http://www.benediktdiemer.com/code/colossus/}} Jupyter Notebook,\footnote{\url{https://jupyter.org}} and Visual Studio Code\footnote{\url{https://github.com/microsoft/vscode}}.

\bibliography{references,references_new}

\end{document}